\documentclass[12pt,a4j]{article}
\usepackage[dvips]{graphicx}
\usepackage{enumerate}
\usepackage{amsmath}
\usepackage{amsfonts}
\usepackage{mathrsfs}

\begin{document}
\baselineskip=5.5mm
\centerline{\bf A direct approach for solving the cubic Szeg\"o equation  }\par
\bigskip
\centerline{Yoshimasa Matsuno\footnote{{\it E-mail address}: matsuno@yamaguchi-u.ac.jp}}\par

\centerline{\it Division of Applied Mathematical Science,}\par
\centerline{\it Graduate School of Science and Technology for Innovation} \par
\centerline{\it Yamaguchi University, Ube, Yamaguchi 755-8611, Japan} \par
\bigskip
\bigskip
\leftline{\bf Abstract}\par
\noindent  We study the cubic Szeg\"o equation which is an integrable nonlinear  non-dispersive and nonlocal evolution equation. In particular, we
present a  direct approach for obtaining the multiphase and multisoliton solutions as well as a special class of periodic solutions.
 Our method is substantially different from the
existing one which relies mainly on the spectral analysis  of the Hankel operator. 
We show that the cubic Szeg\"o equation can be bilinearized through appropriate dependent variable transformations and then
 the solutions  satisfy a set of bilinear equations. The proof is carried out within the framework of an elementary theory of determinants.
Furthermore, we  demonstrate that the eigenfunctions associated with  the multiphase solutions satisfy the Lax pair for  the cubic Szeg\"o equation,
providing an alternative proof of the solutions.
Last, the eigenvalue problem for a periodic solution is solved exactly to obtain the analytical expressions of the eigenvalues. 
 \par

\newpage
\leftline{\bf  1 INTRODUCTION} \par
\bigskip
\noindent {\bf 1.1 Cubic Szeg\"o equation} \par
\medskip
\noindent The cubic Szeg\"o equation has been introduced as a non-dispersive Hamiltonian equation which is reduced from a nonlocal
cubic Schr\"odinger equation called the half-wave equation.$^1$  It may be written in the form
$${\rm i}\,u_t=\Pi_+(|u|^2u), \eqno(1)$$
where $u=u(x, t)$ is a complex function analytic in the upper-half complex  plane, the operator $\Pi_+$ is the Szeg\"o projector defined
by $\Pi_+(\sum_{n=-\infty}^\infty \hat u_n\,e^{{\rm i} nx})=\sum_{n=0}^\infty \hat u_n\,e^{{\rm i} nx}$ for $2\pi$-periodic functions, and
$\Pi_+\left({1\over 2\pi}\int_{-\infty}^\infty\hat u(k)\,e^{{\rm i}kx}dk\right)={1\over 2\pi}\int_{0}^\infty\hat u(k)\,e^{{\rm i}kx}dk$ for functions rapidly vanishing at infinity.
In the latter case, the operator $\Pi_+$ is identified with the orthogonal projector $P_+={1\over 2}(1-{\rm i}H)$, where $H$ is the Hilbert transform 
with the property $He^{{\rm i}kx}={\rm i}\,{\rm sgn}(k)\,e^{{\rm i}kx}$.
The subscript $t$ appended to $u$ denotes partial differentiation.  The most remarkable
feature of the cubic Szeg\"o equation is the complete integrability. Actually, it exhibits the Lax pair structure$^1$
$$H_u\psi=\lambda \psi, \eqno(2a)$$
$$\psi_t=B_u\psi\equiv \left({{\rm i}\over 2}\,H_{u}^2-{\rm i}T_{|u|^2}\right)\psi. \eqno(2b)$$
Here,  $H_u\psi=\Pi_+(u\psi^*)$ is the Hankel operator,  $T_{|u|^2}\psi=\Pi_+(|u|^2\psi)$ is the Toeplitz operator, $\lambda$ is the spectral parameter 
and  $\psi^*$ is the complex conjugate of $\psi$. 
A direct consequence of the Lax pair is the existence  of an infinite number of conservations laws. Among them, the mass $Q$, the momentum $M$ and the energy $E$ defined on $2\pi$-periodic functions
 $$Q={1\over 2\pi}\int_0^{2\pi}|u|^2dx, \quad M={{\rm i}\over 2\pi}\int_0^{2\pi}(u^*u_x-uu_x^*)dx, \quad E={1\over 2\pi}\int_0^{2\pi}|u|^4dx, \eqno(3)$$
are the  fundamental quantities. In view of the integrability property, the analysis of Equation (1) has been performed focusing mainly on the global existence of smooth solutions, the existence of
low regularity solutions and the growth of high Sobolev norms and so on.$^{1-8}$ \par
The construction of the exact solutions such as the soliton  and periodic solutions is another important issue in the soliton theory. There exist several exact methods of solution which include
the inverse scattering transform (IST) method,$^{9-12}$ B\"acklund transformation$^{13-15}$ and direct method (or bilinear transformation method).$^{16-18}$  An approach similar to the IST has been developed  to obtain the
explicit formulas for  solutions to the Cauchy problems of Equation (1) under both the periodic and non-periodic boundary conditions.$^{2, 5, 7, 8}$ 
In the process, the direct and inverse spectral problems of the Hankel operator  played a central role which appears in the Lax pair (2). It turns out, however that
the knowledge of the functional analysis is inevitable to understand the contents of the existing literatures dealing with Equation (1). It will be therefore worthwhile  
to provide an alternative proof of the solutions without recourse to the spectral analysis. \par
The main purpose of the present paper is to derive the multiphase and multisoliton solutions
of Equation (1) by means of the direct method. In so doing, only an elementary theory of determinants is employed. We recall that a similar method was used recently to construct the multiphase solutions 
of a  nonlocal nonlinear Schr\"odinger equation with  focusing nonlinearity$^{19}$ 
$${\rm i}u_t=u_{xx}-2{\rm i}u\Pi_+(|u|^2)_x. $$
Another method of solution was developed which depends on the Lax pair structure of the equation.$^{20}$ \par
\bigskip
\noindent{\bf 1.2 Outline of the paper}\par
  \medskip
\noindent In the remaining part of this section, we summarize the notations. In Section 2, we first bilinearize Equation (1) through appropriate dependent variable transformation to obtain a set
of bilinear equations for the two fundamental tau-functions $f$ and $g$.  The $N$-phase solution ($N$: positive integer) 
can then be expressed in the form $u=g/f$, where both tau-functions have the determinantal structures.  We show that $f$ and $g$ satisfy the bilinear equations by employing various formulas for
determinants in which Jacobi's formula plays a central role. Next, a special class of periodic solutions of the form $ u=Q_m(z)/P_n(z)\ (z=e^{{\rm i}kx}, k>0)$ is presented, where
$P_n(z)$ and $Q_m(z)$ are the $n$th-order and $m$th-order polynomials of $z$, respectively with $0\leq m <n\leq N$.
In Section 3, we work on the eigenvalue problems of the Lax pair (2). In particular, we give a direct proof that the eigenfunctions associated with the $N$-phase solution satisfy (2).
As for the periodic solutions mentioned above, we develop a method for calculating the eigenvalues and provide an explicit example of the traveling wave solution 
which gives rise to  the analytical expressions of the eigenvalues.
In Section 4, we first derive the $N$-soliton solution of Equation (1) by taking a long-wave limit of the $N$-phase solution.  We then develop a direct proof of the $N$-soliton solution. 
Subsequently, the invertibility of the matrix associated with the tau-function $f$ is demonstrated by a simple argument.
Last, the large time asymptotics of the 
solution is briefly described, showing that  no phase shifts appear after the collisions of solitons.
In Section 5, we address concluding remarks. Most of the technical details are explained in appendices.
In Appendix A, after summarizing the basic formulas of determinants, we verify   Lemma 1 in which  various determinantal formulas are established.
In Appendix B, we give the proof of Lemma 2. Appendix C is concerned with the  proof of Lemma 3. 
\par
\bigskip
\noindent{\bf 1.3 Notations}\par
\medskip
 1)  Row vectors \par
$${\bf a}=(a_1, a_2, ..., a_N)=(a_j)_{1\leq j\leq N}=(a_j), \quad {\bf b}=(b_1, b_2, ..., b_N)=(b_j)_{1\leq j\leq N}=(b_j),\quad$$
$$ {\bf c}=(c_1, c_2, ..., c_N)=(c_j)_{1\leq j\leq N}=(c_j), \quad  \quad {\bf d}=(d_1, d_2, ..., d_N)=(d_j)_{1\leq j\leq N}=(d_j),\quad$$
$$ {\bf 1}=\underbrace{(1, 1, ..., 1)}_N=(1)_{1\leq j\leq N}=(1), \eqno(4)$$
where $a_j, b_j, c_j, d_j\ (j=1, 2, ..., N)\in \mathbb{C}$.\par
\medskip
 2) Matrices and cofactors \par
$$F=(f_{jk})_{1\leq j ,k\leq N}\in \mathbb{C}^{N\times N}, \quad F((a_k); (b_j))=\begin{pmatrix} F & {\bf b}^T\\ {\bf a} & 0\end{pmatrix}\in \mathbb{C}^{(N+1)\times(N+1)}, $$
$$ F((a_k), (b_k) ; (c_j), (d_j))=\begin{pmatrix} F & {\bf c}^T & {\bf d}^T\\ {\bf a} & 0 &0 \\ {\bf b} & 0 & 0\end{pmatrix}\in \mathbb{C}^{(N+2)\times(N+2)}\ ({\rm bordered\ matrices}),\eqno(5) $$ 
$$|F|={\rm det}\,F, \quad F_{jk}=\partial |F|/\partial f_{jk}\ ({\rm first\ cofactor\ of}\ f_{jk}), $$
$$ F_{jk, lm}=\partial^2|F|/\partial f_{jl}\partial f_{km},\quad F_{jk,lm}=-F_{kj,lm}=-F_{jk,ml}\ ({\rm second\ cofactor}), \eqno(6)$$
where $f_{jk}\ (j, k=1, 2, ..., N)\in \mathbb{C}$ and the symbol $T$ denotes transpose. \par
\medskip
 3) Bilinear operator \par
$$D_tg\cdot f=g_tf-gf_t. \eqno(7)$$
\par
\bigskip
\noindent {\bf 2 CONSTRUCTION OF MULTIPHASE SOLUTIONS} \par
\medskip
\noindent{\bf 2.1 Bilinearization}\par
\medskip
\noindent In this section, we provide a new explicit $N$-phase solution of Equation (1).  First, according to the prescription of the direct method,$^{16-18}$
 we perform the bilinearization of Equation (1). \par
 \medskip
 \noindent {\bf Theorem 1.} {\it Let the tau-functions $f=f(x, t)$ and $g=g(x,t)$ be polynomials of the variables $z_j=e^{{\rm i}k_jx}\ (j=1,2, ..., N)$, where $k_j$ are arbitrary positive parameters
 sometimes called the wavenumbers.
 The tau-function $f$ is assumed to have no zeros in the upper-half complex plane ${\rm Im}\, x\geq 0$ and approach a constant as ${\rm Im}\, x \rightarrow +\infty$.
 By means of the dependent variable transformations
 $$u={g\over f}, \eqno(8)$$
 $$|u|^2=c+{\rm i}\,{\partial\over \partial t}\,{\rm ln}\,{f^*\over f}, \eqno(9)$$
 Equation (1) is transformed to the set of bilinear equations 
 $${\rm i}D_tf^*\cdot f+cf^*f=g^*g, \eqno(10)$$
 $$D_tf^*\cdot g-{\rm i}cf^*g=h^*f. \eqno(11)$$
 Here, the tau-function $h$ is a polynomial of $z_j\ (j=1, 2, ..., N)$ satisfying the relation $\Pi_+(h/f)^*=0$ and $c$ is a real constant.} \par
 \medskip
 \noindent {\it Proof.} The bilinear equation (10) follows simply by introducing (8) into (9).  It follows from the analyticity of $u$ in the upper-half complex plane 
 that $u_t =\Pi_+(u_t)$, with which Equation (1) can be put into the form $\Pi_+(-u_t-{\rm i}|u|^2u)=0$. If we substitute (8) and (9) into this equation,  we obtain
 $$\Pi_+\left\{{1\over f^*f}(f_t^*g-f^*g_t-{\rm i}cf^*g)\right\}=0. $$
  In view of the relation $\Pi_+(h/f)^*= 0$, the above equation is satisfied identically by virtue of (11).  
 \hspace{\fill}$\Box$ \par
 \medskip
 The explicit form of $h$ given later in the proof of (11) reveals that
the function $(h/f)^*$ is analytic in the lower-half complex plane and satisfies the boundary condition $(h/f)^*=0$ as ${\rm Im}\,x\rightarrow -\infty$, which gives $\Pi_+(h/f)^*= 0$. \par
\medskip
\noindent {\bf 2.2 $N$-phase solution} \par
\medskip
\noindent {\bf Theorem 2.} {\it The $N$-phase solution of Equation (1) admits a determinantal expression in terms of
 the tau-functions $f=f(x, t)$ and $g=g(x, t)$ 
$$u={g\over f}, \quad f=|F|, \quad g=|G|,\eqno(12)$$
where $F$ and $G$ are  $N\times N$ and $(N+1)\times (N+1)$ matrices whose elements are given respectively by
$$F=(f_{jk})_{1\leq j,k\leq N},\quad f_{jk}={\lambda_j\over \lambda_j^2-\mu_k^2}-{\mu_kz_k\over \lambda_j^2-\mu_k^2}\,e^{-{\rm i}(\phi_j+\theta_k)}, \eqno(13a)$$
$$G=(g_{jk})_{1\leq j,k\leq N+1},\quad g_{jk}=f_{jk},\quad \ g_{j,N+1}=e^{-{\rm i}\phi_j},\quad g_{N+1,k}=-1,\quad (1\leq j,k\leq N),\quad g_{N+1, N+1}=0, \eqno(13b)$$
with
$$z_j=e^{{\rm i}k_jx},\ \phi_j=\lambda_j^2t+\phi_{j0},\ \theta_j=-\mu_j^2t+\theta_{j0}, \ (j=1, 2, ..., N),\ c=\sum_{j=1}^N(\lambda_j^2-\mu_j^2). \eqno(13c)$$
Here, $k_j$ are positive parameters and
 $\phi_{j0}$ and $\theta_{j0}$ are real constants. The real parameters $\lambda_j$ and $\mu_j $  are imposed on the condition
$$\lambda_1>\mu_1>\lambda_2>\mu_2>...>\lambda_N>\mu_N>0. \eqno(14)$$
} \par
\bigskip
\noindent {\bf 2.3 Proof of Theorem 2}\par
\medskip
\noindent The proof of Theorem 2 is established by a sequence of steps. First, we provide formulas associated with the tau-functions $f$ and $g$. \par
\medskip
\noindent {\bf Lemma 1.}\par
$$f^*=\kappa\big\{|F|+|F((1);(1/\lambda_j))|\big\},\quad \kappa=\prod_{j=1}^N\left[{\lambda_je^{{\rm i}(\phi_j+\theta_j)}\over -\mu_jz_j}\right], \eqno(15a)$$
$$f_t=-{\rm i}\left|F((\mu_kz_ke^{-{\rm i}\theta_k});(e^{-{\rm i}\phi_j}))\right|, \eqno(15b)$$
$$f_t^*=-{\rm i}\kappa\left|F((\mu_k^2);(1/\lambda_j))\right|, \eqno(15c)$$
$$g=-\left|F((1);(e^{-{\rm i}\phi_j}))\right|, \eqno(15d)$$
$$g^*=\kappa\left|F((\mu_kz_ke^{-{\rm i}\theta_k});(1/\lambda_j))\right|, \eqno(15e)$$
$$g_t={\rm i}\left|F((1);(\lambda_j^2e^{-{\rm i}\phi_j}))\right|, \eqno(15f)$$
$$\left|F((1),(\mu_kz_ke^{-{\rm i}\theta_k});(e^{-{\rm i}\phi_j}),(1/\lambda_j))\right|=-c|F((1);(1/\lambda_j))|+|F((1);(\lambda_j))|-|F((\mu_k^2);(1/\lambda_j))|, \eqno(15g)$$
$$\left|F((1);(\lambda_j^2e^{-{\rm i}\phi_j}))\right|-c\left|F((1);(e^{-{\rm i}\phi_j}))\right|-\left|F((\mu_k^2);(e^{-{\rm i}\phi_j}))\right|=0. \eqno(15h)$$
The proof of Lemma 1 is given in Appendix A. \par
\medskip
\noindent {\bf 2.3.1  Proof of (10)}\par
\medskip
\noindent We show that the tau-functions $f$ and $g$ from (12) and (13) satisfies the bilinear equation (10). Let $P={\rm i}D_tf^*\cdot f+cf^*f-g^*g$. Substituting (15a)-(15e),  $P$ becomes 
$$P={\rm i}\Bigl[-{\rm i}\kappa\left|F((\mu_k^2);(1/\lambda_j))\right||F|+{\rm i}\kappa\big\{|F|+|F((1);(1/\lambda_j))|\big\}\left|F((\mu_kz_ke^{-{\rm i}\theta_k});(e^{-{\rm i}\phi_j}))\right|\Bigr]$$
$$+c\kappa\Big\{|F|+|F((1);(1/\lambda_j))|\Big\}|F|+\kappa\left|F((\mu_kz_ke^{-{\rm i}\theta_k});(1/\lambda_j))\right|\left|F((1);(e^{-{\rm i}\phi_j}))\right|.$$
By using Jacobi's formula
$$ |F((a_k); (c_j))||F((b_k); (d_j))|-|F((a_k); (d_j))||F((b_k); (c_j))|=|F((a_k), (b_k); (c_j), (d_j))||F|, $$
from (A.3) with $a_k=1, b_k=\mu_kz_k\,e^{-{\rm i}\theta_k}, c_j=e^{-{\rm i}\phi_j}, d_j=1/\lambda_j$, one has
$$\left|F((\mu_kz_ke^{-{\rm i}\theta_k});(1/\lambda_j))\right|\left|F((1);(e^{-{\rm i}\phi_j}))\right|
-|F((1);(1/\lambda_j))|\left|F((\mu_kz_ke^{-{\rm i}\theta_k});(e^{-{\rm i}\phi_j}))\right|$$
$$=\left|F((1),(\mu_kz_ke^{-{\rm i}\theta_k});(e^{-{\rm i}\phi_j}),(1/\lambda_j))\right||F|. $$
Then, $P$ reduces to
$$P=\kappa|F|\Bigl[\left|F((\mu_k^2);(1/\lambda_j))\right|+\left|F((1),(\mu_kz_ke^{-{\rm i}\theta_k});(e^{-{\rm i}\phi_j}),(1/\lambda_j))\right|$$
$$-\left|F((\mu_kz_ke^{-{\rm i}\theta_k});(e^{-{\rm i}\phi_j}))\right|+c\big\{|F|+|F((1);(1/\lambda_j))|\big\}\Bigr]. $$
In view of (15g) and the definition of $f_{jk}$ from (13), the above expression simplifies to
\begin{align}
P&=\kappa|F|\Big\{\left|-F((\mu_kz_ke^{-{\rm i}\theta_k});(e^{-{\rm i}\phi_j}))\right|+|F((1);(\lambda_j))|+c|F|\Big\} \notag\\
 &=\kappa|F|\Big\{-\sum_{j,k=1}^N(\lambda_j-\mu_kz_ke^{-{\rm i}(\phi_j+\theta_k)})F_{jk}+c|F|\Big\} \notag\\
 &=\kappa|F|\Big\{-\sum_{j,k=1}^N(\lambda_j^2-\mu_k^2)f_{jk}F_{jk}+c|F|\Big\} \notag\\
 &=\kappa|F|\Big\{-\sum_{j=1}^N(\lambda_j^2-\mu_j^2)|F|+c|F|\Big\} \notag \\
 &=0, \notag
 \end{align}
where in passing to the second line, Formula (A.2) was used while in passing to the fourth line, Formula (A.4) was applied,
thus completing the proof. \hspace{\fill}$\Box$ \par 
\medskip
\noindent {\bf 2.3.2  Proof of (11)}\par
\medskip
\noindent Let $Q=D_tf^*\cdot g-{\rm i}cf^*g$. Substitution of (15a), (15c),(15d) and (15f) into $Q$ gives
$$Q={\rm i}\kappa\left|F((\mu_k^2);(1/\lambda_j))\right|\left|F((1);(e^{-{\rm i}\phi_j}))\right|$$
$$-{\rm i}\kappa\Big\{|F|+|F((1);(1/\lambda_j))|\Big\}\Big\{\left|F((1);(\lambda_j^2e^{-{\rm i}\phi_j}))\right|-c\left|F((1);(e^{-{\rm i}\phi_j}))\right|\Big\}.$$
Introducing (15h) into the second term in the above expression, Q recasts to 
 $$ Q={\rm i}\kappa\Big\{|F((\mu_k^2);(1/\lambda_j))|\left|F((1);(e^{-{\rm i}\phi_j}))\right|-|F((1);(1/\lambda_j))||F((\mu_k^2);(e^{-{\rm i}\phi_j}))|
-|F((\mu_k^2);(e^{-{\rm i}\phi_j}))||F|\Big\}. $$
 One then applies Jacobi's formula to modify $Q$  into the form
 \begin{align}
  Q&={\rm i}\kappa \Big\{|F((\mu_k^2),(1);(1/\lambda_j),(e^{-{\rm i}\phi_j}))|-|F((\mu_k^2);(e^{-{\rm i}\phi_j}))|\Big\}|F| \notag \\
   &={\rm i}\kappa\begin{vmatrix} F & \left({1\over \lambda_j}\right)^T & (e^{-{\rm i}\phi_j})^T \\ 
      (\mu_k^2) & 0 & 0 \\
      (1) & 1 & 0 \end{vmatrix}|F| \notag \\
   &=-{\rm i}\kappa\begin{vmatrix} \left(f_{jk}-{1\over \lambda_j}\right) & (e^{-{\rm i}\phi_j})^T \\ (\mu_k^2) & 0 \end{vmatrix}|F|. \notag
  \end{align}
It follows from (13) that
$$f_{jk}-{1\over \lambda_j}=-{\mu_kz_k\over \lambda_j}\,e^{-{\rm i}(\phi_j+\theta_k)}f_{jk}^*.$$
After substituting this expression into $Q$, one extracts the factor $e^{-{\rm i}\phi_j}/\lambda_j$ from the $j$th row
 and the factor $\mu_kz_ke^{-{\rm i}\theta_k}$ from the
$k$th column, respectively for $j, k=1, 2, .., N.$ Referring to the definition of $\kappa$ from (15a), $Q$ becomes
$$Q={\rm i}|F((\mu_kz_ke^{-{\rm i}\theta_k});(\lambda_j))|^*|F|.$$
Consequently, one can put $Q=h^*f$ with
$$h=-{\rm i}|F((\mu_kz_ke^{-{\rm i}\theta_k});(\lambda_j))|.$$
The rational function $(h/f)^*$ in $e^{-{\rm i}k_jx},\ (j=1, 2, ..., N)$ is analytic in ${\rm Im}\,x\leq 0$ and satisfies the boundary condition
$(h/f)^*=0$ as ${\rm Im}\,x\rightarrow -\infty$. It turns out that $\Pi_+(h/f)^*=0$, as required in
Theorem 1. This completes the proof of (11). \hspace{\fill}$\Box$ \par
\medskip
\noindent {\it Remark }1. \ The special case of the $N$-phase solution with $k_j=1\ (j=1, 2, ..., N)$ has been given in Ref. 7, where an elementary proof of  the invertibility of the
matrix $F$ has been exhibited.  The similar argument will be applied to the matrix $F$ given by (13) as well to establish its invertibility. 
Quite recently, an explicit formula was constructed for the solution of the cubic Szeg\"o equation and its flow was extended to the whole Hardy space $H^2(\mathbb{R})$.$^{8}$ 
The formula established, however depends heavily on the Lax pair structure of the equation.  On the other hand, our method is more direct. Actually, it does not employ the complete integrability of the equation.
The set of rational functions of the form $u=A(z)/B(z)$ has been introduced in Ref. 1, where $A(z)$ and $B(z)$ are polynomials in $z$ whose degrees are $m$ and $n$\ $(0\leq m<n\leq N)$, respectively 
with a condition $B(0)=1$.
This class of solutions would be reduced from the $N$-phase solution by specifying the parameters $\lambda_j$ and $\mu_j$. A few examples will be considered later in subsection 2.4.
\par
\medskip
\noindent{\bf 2.4 Special class of periodic solutions} \par
\medskip
\noindent Here, we present the periodic solutions of Equation (1) of the form
$$u={Q_m(z)\over P_n(z)}, \quad z=e^{{\rm i}kx}\ (k>0), \eqno(16)$$
where $P_n$ and $Q_m$ are polynomials in $z$ whose degrees  are $n$ and $m$ ($0\leq m<n$), respectively, and have no common factors.
In addition, $P(0)=1$ and $P_n$ has no zeros in $|z|\leq 1$, or equivalently ${\rm Im}\,x\geq 0$ to assure the analyticity of $u$.
Below, we give the two examples of  solutions and show that the associated tau-functions satisfy the bilinear equations (10) and (11). \par
\medskip
\noindent {\bf 2.4.1 \ Example 1. $n=N,\ m=N-1$} \par
\medskip
\noindent{\bf Proposition 1.}\ {\it The following tau-functions satisfy the bilinear equations (10) and (11)
$$f=r(z)+r^*(z)f_Nz^N, \quad g=g_0r(z),\quad  h=-{\rm i}\,{c\over 1-\alpha^2}\,g_0^*r^*(z)f_Nz^N, \eqno(17a)$$
with
$$r(z)=\sum_{j=0}^{N-1}a_jz^j, \quad a_0=1,\quad a_j\in \mathbb{C},\ (j=1, 2, ..., N-1), \eqno(17b)$$
$$f_N=\alpha\,e^{-{\rm i}(ct+\phi)},\quad g_0=\sqrt{c(1-\alpha^2)}\,e^{-{\rm i}\left({ct\over 1-\alpha^2}+\delta\right)}
 \ (0<\alpha<1,\ c>0,\ \phi,\delta \in \mathbb{R}). \eqno(17c)$$}
\par
\medskip
\noindent{\it Proof.}\ Direct computations give
$${\rm i}D_tf^*\cdot f+cf^*f=c(1-f_N^*f_N)r^*r=c(1-\alpha^2)r^*r=g^*g,$$
and
$$D_tf^*\cdot g-{\rm i}cf^*g={\rm i}\,{c\over 1-\alpha^2}\,g_0r\,{f_N^*\over z^N}(r+r^*f_Nz^N)=h^*f,$$
showing that the tau-functions (17) satisfy the bilinear equations (10) and (11).  This completes  the proof of Proposition 1. \hspace{\fill}$\Box$ \par
\medskip
\noindent {\it Remark }2. \ To assure  that $f$ has no zeros in $|z|\leq 1$, one must impose the condition on the parameters $a_j\ (j=2, ..., N-1)$.
Since $\lim_{z\rightarrow 0}f=1$, $f$ has no zeros in a small disk centered at $z=0$ (or equivalently, ${\rm Im}\, x>0$). This information will be used to establish the above statement.
If the analytical requirement for $f$ holds true, then the condition $\Pi_+(h/f)^*=0$ imposed in deriving the bilinear equation (11) is
shown to be satisfied. Actually, in view of the facts that $h/f$ is analytic in ${\rm Im}\, x\geq 0$ and  has the boundary value $h/f=0$ as ${\rm Im}\, x\rightarrow +\infty$,
the  condition mentioned above follows immediately. \par
\medskip
\noindent {\it Remark }3. \ We infer that the tau-functions (17) would be reduced  by taking an  appropriate limit of the corresponding tau-functions for the $N$-phase solution given in Theorem 2.
Although the proof of this statement still remains open for general $N$, the reduction procedure can be  performed easily for $N=2$. 
We summarize it shortly. Let $\mu_1=\lambda_2+\epsilon$ and $\mu_2=\lambda_2-\epsilon$ and then take the limit $\epsilon\rightarrow +0$ in the two-phase solution (12) with $k_1=k_2=k$.
This yields the solution of the form $u= g/f$ with
$$f=1-{1\over 2}\left({\lambda_2\over \lambda_1}\,e^{-{\rm i}\{(\lambda_1^2-\lambda_2^2)t+\phi_{10}\}}+e^{-{\rm i}\phi_{20}}\right)\left(e^{-{\rm i}\theta_{10}}+e^{-{\rm i}\theta_{20}}\right)z$$
$$+{\lambda_2\over \lambda_1}\, e^{-{\rm i}\{(\lambda_1^2-\lambda_2^2)t+\phi_{10}+\phi_{20}+\theta_{10}+\theta_{20}\}}z^2\quad (z=e^{{\rm i}kx}), \eqno(18a)$$
$$g={\lambda_1^2-\lambda_2^2\over \lambda_1}\,e^{-{\rm i}(\lambda_1^2t+\phi_{10})}\left\{1-{1\over 2}\,e^{-{\rm i}\phi_{20}}\left(e^{-{\rm i}\theta_{10}}+e^{-{\rm i}\theta_{20}}\right)z\right\}. \eqno(18b)$$
We rewrite the solution in terms of the parameters $a_1$ and $f_2$ given respectively by
$$a_1=-{1\over 2}\,e^{-{\rm i}\phi_{20}}\left(e^{-{\rm i}\theta_{10}}+e^{-{\rm i}\theta_{20}}\right), 
\quad f_2={\lambda_2\over \lambda_1}\,e^{-{\rm i}\{(\lambda_1^2-\lambda_2^2)t+\phi_{10}+\phi_{20}+\theta_{10}+\theta_{20}\}},$$
to obtain
$$u={\lambda_1^2-\lambda_2^2\over \lambda_1}\,e^{-{\rm i}(\lambda_1^2t+\phi_{10})}{1+a_1z\over 1+(a_1^*f_2+a_1)z+f_2z^2}. \eqno(19)$$
Furthermore, if we introduce the new real parameters $\alpha, c, \phi$ and $\delta$ according to the relations
$$\alpha={\lambda_2\over \lambda_1},\quad c=\lambda_1^2-\lambda_2^2, \quad \phi=\phi_{10}+\phi_{20}+\theta_{10}+\theta_{20}, \quad  \delta=\phi_{10}, $$
$f_2$ and $g_0$ can be put into the form 
$$f_2=\alpha\,e^{-{\rm i}(ct+\phi)}, \quad  g_0=\sqrt{c(1-\alpha^2)}\,e^{-{\rm i}\left({c\over 1-\alpha^2}t+\delta\right)}. $$
Plugging these expressions into (19), we can see that the associated tau-functions coincide with those of (17)  with $N=2$. \par
\medskip
\noindent {\bf 2.4.2 \ Example 2. $n=N,\ m=l\ (0\leq l\leq N-1)$} \par
\medskip
\noindent{\bf Proposition 2.}\ {\it The following tau-functions satisfy the bilinear equations (10) and (11)
$$f=1-p^N\xi^N,\quad g={\tilde\alpha} e^{-{\rm i}\omega t}\xi^l, \quad
h={{\rm i}{\tilde\alpha}^*|{\tilde\alpha}|^2p^N\over \{1-(p^*p)^N\}^2}e^{{\rm i}\omega t}\xi^{N-l},\quad \xi=e^{{\rm i}k(x-\tilde ct-x_0)} \eqno(20a)$$
where
$$\tilde c={|{\tilde\alpha}|^2\over kN}{1\over 1-(p^*p)^N}, \quad \omega={|{\tilde\alpha}|^2\over N}\,{N-\{1-(p^*p)^N\}l\over \{1-(p^*p)^N\}^2},
\quad  p, \ {\tilde\alpha}\in \mathbb{C},\  x_0 \in \mathbb{R}, \ 0<|p|<1. \eqno(20b)$$}
\medskip
\noindent{\it Proof.}\ We require that the tau-functions (20) satisfy Equation(10), giving 
$$(k\tilde cN-c)(p^N\xi^N+{p^*}^N\xi^{-N})-2k\tilde cN(p^*p)^N+c\{1-(p^*p)^N\}=|{\tilde\alpha}|^2.$$
Since this relation  must hold for arbitrary $\xi$, we obtain
$$c=k\tilde cN,\quad -2k\tilde cN(p^*p)^N+c\{1+(p^*p)^N\}=|{\tilde\alpha}|^2, \eqno(20c)$$
from which the expression of $\tilde c$ given in (20b) follows. \par
On the other hand, taking into account (20c), Equation (11) reduces to
$$-{\rm i}\tilde\alpha e^{-{\rm i}\omega t} p^{-N}\xi^{-(N-l)}\Big\{\{(p^*p)^N-1\}(\omega +lk\tilde c)+Nk\tilde c+(\omega -Nk\tilde c+lk\tilde c)f\Big\}=h^*f. $$
Comparing the coefficients of $f$ on both sides, one has 
$$\{(p^*p)^N-1\}(\omega +lk\tilde c)+Nk\tilde c=0, \quad h^*=-{\rm i}\tilde\alpha e^{-{\rm i}\omega t} p^{-N}(\omega -Nk\tilde c+lk\tilde c) \xi^{-(N-l)}. $$
By a straightforward computation using the above two relations and $\tilde c$ from (20b), the expressions of $h$ and $\omega$ 
are obtained as indicated by (20a) and (20b), respectively.  Last, the condition $\Pi_+(h/f)^*=0$   is found to be satisfied since the rational
function $h/f$ is analytic in ${\rm Im}\, x\geq0$ and has the boundary value $h/f=0$ as ${\rm Im}\, x\rightarrow +\infty$. \hspace{\fill}$\Box$ \par
\medskip
We note that the solution of the traveling wave type given in Proposition 2 has been presented in Ref. 1. Its proof is, however based on a lengthy spectral analysis of the
Hankel operator. In Section 3, we will solve exactly the eigenvalue problem (2a) for the special case  of the  solution (17) with $r(z)=1$. \par
\bigskip
\noindent {\bf 3 EIGENVALUE PROBLEMS} \par
\medskip
\noindent{\bf 3.1 Eigenvalue problem associated with the $N$-phase solution} \par
\medskip
\noindent {\bf 3.1.1 Spatial part of the Lax pair}\par
\medskip
\noindent Here, we consider the spatial part of the  Lax pair (2a) for the discrete eigenvalues. To be more specific, it reads
$$H_u\psi_j=\hat\lambda_j\psi_j, \quad (j=1, 2, ..., N), \eqno(21)$$
where $\hat\lambda_j \in \mathbb{C}$ are eigenvalues and $\psi_j$ are corresponding eigenfunctions. We show that Equation (21) can be solved explicitly
for the $N$-phase solution. Firstly, we put $\hat\lambda_j=\lambda_j\,e^{{\rm i}\chi_j}$, where $\lambda_j$ are defined in (14) and $\chi_j$ are  constant parameters
and introduce the new variables $u_j$ according to the relations $\psi_j=e^{{\rm i\over 2}(\phi_j-\chi_j)}u_j$ with the parameters $\phi_j$ being given in (13c).
Since  $H_u$ is a $\mathbb{C}$-antilinear operator,
Equation (21) is  rewritten in the form
$$H_uu_j=\lambda_je^{{\rm i}\phi_j}u_j, \quad (j=1, 2, ..., N). \eqno(22)$$
Referring to (15d) for the tau-function $g$, the $N$-phase solution $u$ can be decomposed in terms of $u_j$ as
$$u=\sum_{j=1}^Nu_j, \quad u_j={g_j\over f}, \quad g_j=e^{-{\rm i}\phi_j}\sum_{j=1}^NF_{jk}, \quad (j=1, 2, ..., N). \eqno(23)$$
We substitute  $u_j$ from (23) into (22) and recast it into the form
$$\Pi_+\Big\{{1\over f^*f}\left(gg_j^*-\lambda_je^{{\rm i}\phi_j}g_jf^*\right)\Big\}=0. \eqno(24)$$
Thus, the original eigenvalue problem (21) has been transformed to solving Equation (24).
\par
Now, we define the  variable $\hat g$ by
$$\hat g=\sum_{j=1}^Ng_js^{j-1}=
\sum_{j,k=1}^NF_{jk}e^{-{\rm i}\phi_j}s^{j-1}=-|F((1);(e^{-{\rm i}\phi_j}s^{j-1}))|. \eqno(25)$$
Multiplying (24) by $s^{j-1}\ (s\in \mathbb{R})$ and summing up with respect to $j$ from 1 to $N$, the eigenvalue problem under consideration
is rephrased as follows.\par
\medskip
\noindent {\bf Proposition 3.}\ {\it The eigenvalue problem (21) is found to admit the exact solutions $\psi_j=e^{{\rm i\over 2}(\phi_j-\chi_j)}{g_j\over f}\ (j=1, 2, ..., N)$,
if one could verify the equation
$$\Pi_+\left\{{1\over f^*f}\left(g{\hat g}^*-\sum_{j=1}^N\lambda_je^{{\rm i}\phi_j}g_js^{j-1}f^*\right)\right\}=0. \eqno(26)$$ 
By setting the coefficient of $s^{j-1}$ zero on the left-hand side of (26), Equation (24) follows immediately.} \par
\medskip
 To proceed, we prepare the following lemma which makes the proof of Proposition 3 clear. \par
\medskip
\noindent{\bf Lemma 2.} \par
$${\hat g}^*=\kappa\left|\hat F((\mu_kz_ke^{-{\rm i}\theta_k});(s^{j-1}/\lambda_j))\right|, \quad \hat F=(\hat f_{jk})_{1\leq j,k\leq N}, \quad  \hat f_{jk}= f_{jk}-{1\over \lambda_j}, \eqno(27a)$$
$$\sum_{j=1}^N\lambda_je^{{\rm i}\phi_j}g_js^{j-1}=-|\hat F((1);(\lambda_js^{j-1}))|, \eqno(27b)$$
$$f^*=\kappa|\hat F|, \eqno(27c)$$
$$\left|\hat F((\mu_kz_ke^{-{\rm i}\theta_k});(e^{-{\rm i}\phi_j}))\right|=|\hat F((\mu_k^2);(1/\lambda_j))|+c|\hat F|, \eqno(27d)$$
$$\left|\hat F((1),(\mu_kz_ke^{-{\rm i}\theta_k});(e^{-{\rm i}\phi_j}),(s^{j-1}/\lambda_j))\right|$$
$$=-c|\hat F((1);(s^{j-1}/\lambda_j))|+|\hat F((1);(\lambda_js^{j-1}))|-|F((\mu_k^2);(s^{j-1}/\lambda_j))|, \eqno(27e)$$
$$|\hat F|=|F|+|\hat F((1);(1/\lambda_j))|. \eqno(27f)$$
\par
\noindent The proof of Lemma 2 is given in Appendix B. \par
\medskip
\noindent{\it Proof of Proposition 3.}\ Let $R=g{\hat g}^*-\sum_{j=1}^N\lambda_je^{{\rm i}\phi_j}g_js^{j-1}f^*.$
Referring to (15d),(27a)-(27c),  $R$ reduces  to
$$R=-\kappa |\hat F((1);(e^{-{\rm i}\phi_j}))|\left|\hat F((\mu_kz_ke^{-{\rm i}\theta_k});(s^{j-1}/\lambda_j))\right|+\kappa |\hat F((1);(\lambda_js^{j-1}))||\hat F|. $$
If one applies Jacobi's formula to the first term and then introduces (27d) into the resultant expression, $R$ becomes
$$R=-\kappa\Big\{\left|\hat F((1),(\mu_kz_ke^{-{\rm i}\theta_k});(e^{-{\rm i}\phi_j}),(s^{j-1}/\lambda_j))\right|+c|\hat F((1);(s^{j-1}/\lambda_j))|-|\hat F((1);(\lambda_js^{j-1}))|\Big\}|\hat F|$$
$$-\kappa |\hat F((1);(s^{j-1}/\lambda_j))||\hat F((\mu_k^2);(1/\lambda_j))|. $$
The above expression of $R$ is further modified by introducing (27e) into the first term and applying Jacobi's formula to the last term, giving
$$R=\kappa\Big\{|\hat F((\mu_k^2);(s^{j-1}/\lambda_j))||\hat F|-|\hat F((1);(s^{j-1}/\lambda_j))||\hat F((\mu_k^2);(s^{j-1}/\lambda_j))|\Big\}. $$
Using (27f) and the relation
$$\hat f_{jk}=-{\mu_kz_k\over \lambda_j}\,e^{-{\rm i}(\phi_j+\theta_k)}f_{jk}^*, \eqno(28)$$
which follows from (13), $R$ reduces to
$$R=\kappa\begin{vmatrix} \left(-{\mu_kz_k\over \lambda_j}\,e^{-{\rm i}(\phi_j+\theta_k)}f_{jk}^*\right) & (s^{j-1}/\lambda_j)^T \\ (\mu_k^2) & 0 \end{vmatrix}|F|. $$
By extracting the factor $e^{-{\rm i}\phi_j}/\lambda_j$ from the $j$th row and the factor $\mu_kz_k\,e^{-{\rm i}\theta_k}$ from the $k$th column, respectively 
for $j, k=1, 2, ..., N$ and taking into account the definition of $\kappa$
from (15a), one finally arrives at the expression of $R$
$$R=-|F((\mu_kz_ke^{-{\rm i}\theta_k});(e^{-{\rm i}\phi_j}s^{j-1}))|^*|F|. $$
Substitution of the above expression into Equation (26) recasts it into the form 
$$\Pi_+\left\{{1\over f}\left|F\left((\mu_kz_ke^{-{\rm i}\theta_k}); (e^{-{\rm i}\phi_j}s^{j-1})\right)\right|\right\}^*=0.$$
Observe that the quantity in the parentheses is analytic in ${\rm Im}\,x\geq 0$ and vanishes as ${\rm Im}\,x\rightarrow +\infty$. Hence, in view of the property of 
 the operator $\Pi_+$, the above
equation is satisfied automatically. \hspace{\fill}$\Box$ \par
\medskip
\noindent  {\bf 3.1.2 Temporal part of the Lax pair}\par
\medskip
\noindent The temporal part of the Lax pair (2b) reads
$$\psi_{j,t}= \left({{\rm i}\over 2}\,H_{u}^2-{\rm i}T_{|u|^2}\right)\psi_j, \quad (j=1, 2, ..., N). \eqno(29)$$
We operate $H_u$ on (21) and use the $\mathbb{C}$-antilinear property of $H_u$ to obtain $H_u^2\psi_j=\lambda_j^2\psi_j$. 
Inserting this equation into (29) and noting that $T_{|u|^2}$ is a $\mathbb{C}$-linear operator, Equation (29) becomes
$$u_{j,t}=-{\rm i\,}T_{|u|^2}u_j=-{\rm i}\,\Pi_+(|u|^2u_j). \eqno(30)$$
If we multiply $s^{j-1}$ and sum with respect to $j$ and take into account the relation $\Pi_+(\hat g/f)=\hat g/f$,  Equation (30) can be put into the form
$$\Pi_+\left\{\left({\hat g\over f}\right)_t+{\rm i}\,|u|^2\,{\hat g\over f}\right\}=0, \eqno(31)$$
where $\hat g$ is defined by (25). Substituting (9) into (31), we finally obtain the equation which is given in the following proposition.\par
\medskip
\noindent{\bf Proposition 4.}{\it\ The eigenvalue problem (29) admits the exact solutions $\psi_j=e^{{\rm i\over 2}(\phi_j-\sigma_j)}\,
{g_j\over f},\ (j=1, 2, ..., N)$,
provided that the equation
$$\Pi_+\left\{{1\over f^*f}\Big(\hat g_tf^*+{\rm i}c\hat gf^*-{f_t}^*\hat g\Big)\right\}=0, \eqno(32)$$
holds.}
\par
\medskip
\noindent{\it Proof.}\ Let $S=\hat g_tf^*+{\rm i}c\hat gf^*-{f_t}^*\hat g$.  Differentiation of $\hat g$ from the last expression of (25) by $t$ gives
$$\hat g_t={\rm i}\left|F((1),(\mu_kz_ke^{-{\rm i}\theta_k});(s^{j-1}e^{-{\rm i}\phi_j}),(e^{-{\rm i}\phi_j}))\right|+{\rm i}|F((1);(\lambda_j^2s^{j-1}e^{-{\rm i}\phi_j}))|. $$
Substituting $f^*$ from (15a), ${f_t}^*$ from (15c), $\hat g$ from (25) as well as   Formula (cf. (27e))
$$\left|F((1),(\mu_kz_ke^{-{\rm i}\theta_k});(e^{-{\rm i}\phi_j}s^{j-1}),(e^{-{\rm i}\phi_j}))\right|$$
$$=c|F((1);(s^{j-1}e^{-{\rm i}\phi_j}))|+|F((\mu_k^2);(s^{j-1}e^{-{\rm i}\phi_j}))|-|F((1);(\lambda_j^2s^{j-1}e^{-{\rm i}\phi_j}))|, $$
into $S$, one finds that
$$S={\rm i}\kappa\Big\{|F((\mu_k^2);(s^{j-1}e^{-{\rm i}\phi_j}))||F|+|F((\mu_k^2);(s^{j-1}e^{-{\rm i}\phi_j}))|F((1);(1/\lambda_j))|$$
$$-|F((\mu_k^2);(1/\lambda_j))||F((1);(s^{j-1}e^{-{\rm i}\phi_j}))|\Big\}.$$
It follows by applying Jacobi's formula to the sum of the second and third terms that
\begin{align}
S&={\rm i}\kappa\Big\{|F((\mu_k^2);(s^{j-1}e^{-{\rm i}\phi_j}))|+|F((\mu_k^2),(1);(s^{j-1}e^{-{\rm i}\phi_j}),(1/\lambda_j)|\Big\}|F| \notag \\
&={\rm i}\kappa|\hat F((\mu_k^2);(s^{j-1}e^{-{\rm i}\phi_j}))||F| \notag \\
&=-{\rm i}\kappa\left|F((\mu_kz_ke^{-{\rm i}\theta_k});(\lambda_js^{j-1}))\right|^*|F|, \notag
\end{align}
where in passing to the last line, an argument developed in proving Proposition 3 has been used (see (28) and a subsequent sentence).
With the above $S$, Equation (32) becomes
$$\Pi_+\Big\{{1\over f} \left|F((\mu_kz_ke^{-{\rm i}\theta_k});(\lambda_js^{j-1}))\right|\Big\}^*=0.$$
Since the quantity in the parentheses is analytic in ${\rm Im}\,x\geq 0$ and vanishes as ${\rm Im}\,x\rightarrow +\infty$, the above
equation is satisfied automatically. \hspace{\fill}$\Box$ \par
\medskip
\noindent{\bf 3.2 Eigenvalue problem associated with periodic solutions (16)}\par
\medskip
\noindent Here, we develop a method for computing the eigenvalues of Equation (2a) for the periodic solutions (16). 
As an example, we consider  a special case of (16) exemplified in (17)
$$u={g\over f}, \quad f=1+f_Nz^N, \quad g=g_0, \eqno(33)$$
where $f_N$ and $g_0$ are given in (17c). 
It is understood that all the computations are carried out at an initial time $t=0$.
We show that the eigenvalue problem for the periodic solution (33) can be solved exactly to obtain the analytical expressions of the eigenvalues. \par
\medskip
\noindent{\bf 3.2.1 General procedure for computing eigenvalues} \par
\medskip
\noindent By employing the partial fraction decomposition, the solution (16) with $n=N$ and $m=N-1$ can be represented in the form
$$u=\sum_{j=1}^N{\alpha_j\over 1-p_jz}, \quad \alpha_j, p_j \in \mathbb{C}, \ 0<|p_j|<1,\quad (j=1, 2, ..., N). \eqno(34)$$
We seek the solution $\psi$ of the form
$$\psi=\sum_{j=1}^N{\beta_j\over 1-p_jz},\quad \beta_j\in \mathbb{C},\quad (j=1, 2, ..., N) . \eqno(35)$$
A simple computation using (34) and (35) gives
$$H_u\psi=-\sum_{j,k=1}^N{\alpha_j{\beta_k}^*\over 1-p_j{p_k}^*}\,\Pi_+\left({{1\over p_j}\over z-{1\over p_j}}-{{p_k}^*\over z-{p_k}^*}\right). \eqno(36a)$$
The relations
$$\Pi_+\left({{1\over p_j}\over z-{1\over p_j}}\right)={{1\over p_j}\over z-{1\over p_j}}, \quad \Pi_+\left({{p_k}^*\over z-{p_k}^*}\right)=0,$$
 stem from the property of the operator $\Pi_+$ and the conditions $0<|p_j|<1\ (j=1, 2, ..., N)$, with which (36a) becomes
$$H_u\psi=\sum_{j,k=1}^N{\alpha_j{\beta_k}^*\over 1-p_j{p_k}^*}{1\over 1-p_jz}. \eqno(36b)$$
If we substitute (35) and (36b) into Equation (2a) and compare the coefficient of $1/(1-p_jz)$
on both sides, we obtain  the system of linear algebraic equations for $\beta_j$ 
$$\sum_{k=1}^N{\alpha_j{\beta_k}^*\over 1-p_j{p_k}^*}=\lambda\beta_j, \quad (j=1, 2, ..., N). \eqno(37)$$
\par
Let us define the $N\times N$ matrix $C$  and the column vector ${\boldsymbol\beta}$ by
$$C=(c_{jk})_{1\leq j,k\leq N}, \quad c_{jk}={\alpha_j\over 1-p_j{p_k}^*}, \quad  {\boldsymbol\beta}=(\beta_1, \beta_2, ..., \beta_N)^T.\eqno(38)$$
Then, the linear system (37) and  its complex conjugate expression  can be written respectively in the form
$$C{\boldsymbol\beta}^*=\lambda{\boldsymbol\beta}, \quad C^*{\boldsymbol\beta}=\lambda^*{\boldsymbol\beta}^*. \eqno(39)$$
It follows from (39) that
$$ CC^*{\boldsymbol\beta}=\lambda^*\lambda{\boldsymbol\beta}. \eqno(40)$$
One can show that Equation (40) is also derived from the equation $H_u^2\psi=\lambda^*\lambda\psi$ by operating $H_u$ on (36b).
In addition to (40), we  consider the eigenvalue problem
$$C{\boldsymbol\gamma}=\lambda{\boldsymbol\gamma}, \eqno(41)$$
where ${\boldsymbol\gamma}=(\gamma_1, \gamma_2, ..., \gamma_N)^T,\ \gamma_j\in \mathbb{C}$\ $(j=1. 2, ..., N)$ is the eigenvector.
The eigenvalues of the above equations are obtained by solving  the characteristic equations 
$$\chi_C(\lambda)\equiv|C-\lambda I_N|=0, \eqno(42)$$
$$ \chi_{CC^*}(\Lambda)\equiv|CC^*-\Lambda I_N|=0, \quad \Lambda= \lambda^*\lambda, \quad (I_N:\ N\times N\ {\rm unit \ matrix}), \eqno(43)$$
where $\chi_C(\lambda)$ and $\chi_{CC^*}(\Lambda)$ are the characteristic polynomials of the matrices $C$ and $CC^*$, respectively. \par
Let $\hat\lambda_j$ and $\Lambda_j\ (j=1, 2, ..., N)$ be the set of eigenvalues of $C$ and $CC^*$, respectively. It then follows that $\prod_{j=1}^N\hat\lambda_j=|C|$
and $\prod_{j=1}^N\Lambda_j=|CC^*|$. Since $|CC^*|=|C||C|^*$, one obtains  $\prod_{j=1}^N\Lambda_j=\prod_{j=1}^N(\hat\lambda_j\hat\lambda_j^*)$. This relation will be 
confirmed later in Remark 4.
\par
\medskip
\noindent{\bf 3.2.2  Computation of the matrix elements }\par
\medskip
\noindent In order to solve Equations (42) and (43), we need to compute the matrix elements of $C$ and $CC^*$.
To this end, we express the coefficients $\alpha_j$ in (38) in terms of $p_j$ whereas the latter are determined uniquely  from the tau-function $f$.
\par
First, we note that the solution (34) takes the form $u=g/f$, where the tau-functions $f$ and $g$ can be written as
$$ f=\prod_{j=1}^N(1-p_jz)=\sum_{j=0}^N(-1)^j\sigma_jz^j, 
\quad       g=\sum_{j=1}^N\alpha_jw_j, \quad  w_j=\prod_{\substack{k=1\\ (k\not= j)}}^N(1-p_kz)=\sum_{n=0}^{N-1}(-1)^n\sigma_{n,j}z^n. \eqno(44a)$$
with $\sigma_j$ being the elementary symmetric polynomials of $p_j$ defined by
$$\sigma_0=1, \quad \sigma_1=\sum_{j=1}^Np_j, \quad \sigma_2=\sum_{1\leq j<k\leq N}p_jp_k, ...,  \quad \sigma_N=\prod_{j=1}^Np_j, \eqno(44b)$$
and
$$\sigma_{n,j}=\sum_{l=0}^n(-1)^lp_j^l\sigma_{n-l}, \quad \sigma_{0,j}=1\ (j=1, 2, ..., N). \eqno(44c)$$
The following proposition provides a method for determining  $\alpha_j$. \par
\medskip
\noindent{\bf Proposition 5.}{\it \ The elements $\alpha_j\ (j=1, 2, ..., N)$ in (38) characterizing the periodic solution (17) 
are given explicitly by
$$\alpha_j={g_0\over |A|}\sum_{k=1}^N(-1)^kA_{kj}a_{k-1}, \quad (j=1, 2, ..., N), \eqno(45)$$
which are obtained by solving the system of linear algebraic equations
$$\sum_{k=1}^Na_{jk}\alpha_k=(-1)^jg_0a_{j-1},\quad (j=1, 2, ..., N). \eqno(46)$$
Here,  $A=(a_{jk})_{1\leq j,k\leq N}$ is the  matrix with elements $a_{jk}=\sum_{l=1}^j(-1)^l\sigma_{j-l}p_k^{l-1}$ and $A_{jk}$ is the cofactor of $a_{jk}$.} \par
\par
\medskip
\noindent{\it Proof.}\ It follows from (44c) that
$$\sum_{j=1}^N\alpha_j\sigma_{n,j}=\sum_{l=0}^n(-1)^l\sigma_{n-l}J_{l}, \quad J_l=\sum_{j=1}^N\alpha_jp_j^l.$$
Substituting this relation into $g$ from (44a) gives
$$g=\sum_{n=0}^{N-1}(-1)^nz^n\sum_{l=0}^n(-1)^l\sigma_{n-l}J_l. \eqno(47)$$
We require that (47) is equal to $g$ from (17) and then compare the coefficient of $z^n$ on both sides, obtaining
 $(-1)^n\sum_{l=0}^n(-1)^l\sigma_{n-l}J_l=g_0a_n$. Substituting the expression of $J_l$ into this relation, we arrive at
 the equations that determine $\alpha_j$
 $$\sum_{k=1}^N\sum_{l=0}^j(-1)^l\sigma_{j-l}p_k^l\alpha_k=(-1)^jg_0a_j, \quad (j=0, 1,  ..., N-1), $$
which becomes (46) by replacing $j$ by $j-1$ and rewriting it in terms of the matrix elements $a_{jk}$. 
The solution of  (46) follows simply by using Cramer's rule as indicated by (45).
 \hspace{\fill}$\Box$ \par
 \medskip
  Next, we derive the expression of $p_j$ in terms of $a_j$ and $f_N$. First of all, we equate $f$ from (44a) with $f$ from (17) to give
  $$\sum_{j=0}^N(-1)^j\sigma_jz^j=1+\sum_{j=1}^{N-1}(a_j+{a_{N-j}}^*f_N)z^j+f_Nz^N. $$
  By comparing the coefficient of $z^j$ on both sides, we find
  $$\sigma_j=(-1)^j(a_j+{a_{N-j}}^*f_N), \ (j=1, 2, ..., N-1), \quad \sigma_N=(-1)^Nf_N. \eqno(48)$$
 The parameters ${p_j}^{-1}$ are obtained by solving the algebraic equation $\sum_{j=0}^N(-1)^j\sigma_jz^j=0$. 
 These informations about $\alpha_j$ and $p_j$ are sufficient to determine the matrix $C$. 
 \par
\bigskip
\noindent{\bf 3.2.3  Example}\par
\medskip
\noindent We compute the eigenvalues associated with Equation (2a) for the periodic solution given by (33).
In this case, $a_j=0$ for $j=1, 2, ..., N-1$. 
It follows from (48) that
$$\sigma_j=0,\quad (j=1, 2, ..., N-1), \eqno(49)$$
 and $p_j$ are roots of the equation $p^N=-f_N$, 
giving
$$p_j=f_N^{1\over N}\,e^{{\pi {\rm i}\over N}(2j+1)}, \quad (j=1, 2, ..., N). \eqno(50)$$
\par
The parameters $\alpha_j$ can be evaluated as follows. In view of (49), the element $a_{jk}$ becomes $(-1)^jp_k^{j-1}$ and hence
$${A_{1j}\over |A|}=-{V_{1j}\over |V|}={(-1)^{N}\prod_{l=1}^Np_l\over p_j^N\prod_{\substack{l=1\\(l\not=j)}}^N\left(1-{p_l\over p_j}\right)}, \quad (j=1, 2, ..., N), $$
where $V=(v_{jk})_{1\leq j,k\leq N}, v_{jk}=p_k^{j-1}$ is the Vandermonde matrix and $V_{1j}$ is the cofactor of the element $v_{1j}$.
Thanks to (49), $\sigma_{n,j}$ from (44c) simplifies to $\sigma_{n,j}=(-1)^np_j^n$. 
  It turns out from (44a) that 
$$\prod_{\substack{l=1\\(l\not=j)}}^N\left(1-{p_l\over p_j}\right)=w_j\left({1\over p_j}\right)=N. $$
Furthermore, referring to the relations $\prod_{l=1}^Np_l=\sigma_N=(-1)^Nf_N$ and $p_j^N=-f_N$ by (50), we  obtain
$A_{1j}/|A|=-1/N$. If we substitute  this relation into (45) with $a_0=1,a_j=0\ (j=1, 2, ..., N-1)$, we finally find the desired expression  $\alpha_j=g_0/N$. 
Thus, with this relation and $p_j$ from (50), the matrix element $c_{jk}$ from (38) takes the simple form
$$c_{jk}={{g_0\over N}\over 1-\alpha^{2\over N}e^{{2\pi{\rm i}\over N}(j-k)}}, \quad (j, k=1, 2, ..., N). \eqno(51)$$ 
\par
The following proposition provides the eigenvalues of the characteristic equation (42) for the  matrix $C$ with $C=(c_{jk})$ being given by  (51). \par
\medskip
\noindent{\bf Proposition 6.}{\it \ The eigenvalue problem (41) associated with the periodic solution (33) has the $N$ simple eigenvalues
$$\hat\lambda_j={g_0\,\alpha^{2(j-1)\over N}\over 1-\alpha^2}, \quad (j=1, 2,  ..., N). \eqno(52)$$}
\medskip
\noindent{\it Proof.}\ Let $C=g_0\tilde C/N$ and $\hat\lambda=g_0\tilde\lambda/N$ with
$$\tilde C=(\tilde c_{jk})_{1\leq j,k\leq N}, \quad \tilde c_{jk}={1\over 1-\beta\zeta^{j-k}}, \quad \beta=\alpha^{2\over N}, \quad \zeta=e^{2\pi{\rm i}\over N}, \eqno(53a)$$
and
$$c_0={1\over 1-\beta}-\tilde\lambda,\quad c_j={1\over 1-\beta\zeta^{-j}}, \quad (j=1, 2, ..., N-1). \eqno(53b)$$
Then, the characteristic polynomial from (42) takes the form
$$\chi_C(\tilde\lambda)=\begin{vmatrix} c_0   & c_1 & c_2    & \cdots& c_{N-1} \\
                                      c_{N-1}  & c_0 & c_1  &\cdots& c_{N-2}\\
                                      \vdots& \vdots & \vdots & \ddots &\vdots\\
                                      c_1& c_2 &c_3 & \cdots& c_0
                                      \end{vmatrix}. \eqno(54a)$$
 Since  $\chi_C(\tilde\lambda)$ is the determinant of the circulant matrix, it can be evaluated explicitly to give$^{21
 }$
 $$\chi_C(\tilde\lambda)=\prod_{j=0}^{N-1}(c_0+\zeta^jc_1+\zeta^{2j}c_2 + ... +\zeta^{(N-1)j}c_{N-1}). \eqno(54b)$$
\par
Now, we compute the quantity $P_j\equiv \sum_{l=0}^{N-1}\zeta^{jl}c_l.$  It follows from (53b) and the relation $\zeta^N=1$ that
\begin{align}
P_j&=c_0+\sum_{l=1}^{N-1}{\zeta^{jl}\over 1-\beta\zeta^{-l}} \notag \\
   &=c_0-{1\over 1-\beta}+\sum_{l=1}^N\sum_{n=0}^\infty\beta^n\zeta^{(j-n)l}. \tag{55}
   \end{align}
   The Taylor expansion in the second line is justified due to the condition $0<\beta<1$. 
   We evaluate  the third term of (55) by splitting the sum with respect to $n$ as
   $$\sum_{n=0}^\infty\beta^n\zeta^{(j-n)l}=\sum_{n=0}^{j-1}\beta^n\zeta^{(j-n)l}+\beta^j+\sum_{n=j+1}^\infty\beta^n\zeta^{(j-n)l}.\eqno(56a) $$
   The first term on the right hand side of (56a) turns out to be
\begin{align}
\sum_{n=j+1}^\infty\beta^n\zeta^{(j-n)l}&= \sum_{n=1}^\infty \beta^{n+j}\zeta^{-nl} \notag \\
                                        &=\sum_{m=1}^\infty \beta^{mN+j}\zeta^{-mNl}+\sum_{\substack{n=1 \\ (n\not= N, 2N, ...)}}^\infty \beta^{n+j}\zeta^{-nl} \notag \\
                                        &={\beta^j\over 1-\beta^N}-\beta^j+\sum_{\substack{n=1 \\(n\not= N, 2N, ...)}}^\infty \beta^{n+j}\zeta^{-nl}. \tag{56b} 
\end{align}
It  follows from (56b) that
\begin{align}
\sum_{l=1}^N\sum_{n=j+1}^\infty\beta^n\zeta^{(j-n)l}
&=\left({\beta^j\over 1-\beta^N}-\beta^j\right)N+\sum_{\substack{n=1 \\(n\not= N, 2N, ...)}}^\infty\beta^{n+j}{\zeta^{-n}(1-\zeta^{-Nn})\over 1-\zeta^{-n}} \notag \\
&=\left({\beta^j\over 1-\beta^N}-\beta^j\right)N. \tag{56c}
\end{align}
Note that the second term of the first line vanishes due to the relation $\zeta^{-Nn}=1$.
Using  (56c), the third term of (55) becomes
\begin{align}
\sum_{l=1}^N\sum_{n=0}^\infty\beta^n\zeta^{(j-n)l} 
&=\sum_{l=1}^N\sum_{n=0}^{j-1}\beta^n\zeta^{(j-n)l}+N\beta^j+\sum_{l=1}^N\sum_{n=j+1}^\infty\beta^n\zeta^{(j-n)l} \notag \\
&=\sum_{n=0}^{j-1}\beta^n\,{\zeta^{j-n}-\zeta^{(j-n)(N+1)}\over 1-\zeta^{j-n}}+N\beta^j+\left({\beta^j\over 1-\beta^N}-\beta^j\right)N \notag \\
&={N\beta^j\over 1-\beta^N}. \tag{56d}
\end{align}
Substituting (56d) into (55), the characteristic polynomial from (54b) is simplified to
$$\chi_C(\tilde\lambda)=\prod_{j=0}^{N-1}\left(-\tilde\lambda+{N\beta^j\over 1-\beta^N}\right). $$
Thus, the characteristic equation $\chi_C(\tilde\lambda)=0$ yields the $N$ simple eigenvalues as shown by (52). \hspace{\fill}$\Box$ \par
\medskip
The eigenvalues of the characteristic equation (43) for the  matrix $CC^*$ can be obtained in the same way, which is stated in the following proposition. \par
\medskip
\noindent{\bf Proposition 7.}{\it\ The eigenvalue problem (40)  associated with  the periodic solution (33) has the multiple eigenvalues
$$\Lambda_1={c\over 1-\alpha^2}, \quad \Lambda_j={c\alpha^2\over 1-\alpha^2},\quad (j=2, 3, ..., N). \eqno(57)$$}
 \par
\medskip
\noindent{\it Proof.}\ We compute the matrix elements $(CC^*)_{jk}$ for $j\not=k$ and $j=k$, separately.
For $j\not=k$, it follows from (51) that 
$$(CC^*)_{jk}={|g_0|^2\over \beta N^2}{1\over \zeta^{j}-\zeta^{k}}\sum_{l=1}^N\left({\beta \zeta^{j}\over 1-\beta \zeta^{j-l}}-{\beta \zeta^{k}\over 1-\beta \zeta^{k-l}}\right). \eqno(58)$$
We  apply the Taylor expansion and use the relation $\zeta^{-Nn}=1$ to obtain
\begin{align}\sum_{l=1}^N{1\over 1-\beta \zeta^{j-l}}
&=\sum_{l=1}^N\sum_{n=0}^\infty\beta^n \zeta^{(j-l)n} \notag \\
&=\sum_{m=0}^\infty\beta^{Nm}\sum_{l=1}^N\,1+\sum_{\substack{n=1\\ (n\not=N, 2N, ...)}}^\infty\beta^n\zeta^{jn}{\zeta^{-n}(1-\zeta^{-Nn})\over 1-\zeta^{-n}} \notag \\
&={N\over 1-\alpha^2}. \tag{59}
\end{align}
Inserting this expression and the corresponding one with $j$ replaced by $k$ into (58) and noting $|g_0|^2=c(1-\alpha^2)$, we arrive at 
$$(CC^*)_{jk}={|g_0|^2\over N}{1\over 1-\alpha^2}={c\over N}. \eqno(60)$$
For $j=k$, on the other hand, we find that
\begin{align}
(CC^*)_{jj}
&={|g_0|^2\over N^2}\sum_{l=1}^N{1\over \left(1-\beta \zeta^{(j-l)}\right)^2} \notag \\
&={|g_0|^2\over N^2}{\partial\over\partial\beta}\sum_{l=1}^N{\zeta^{l-j}\over 1- \beta \zeta^{j-l}} \notag \\
&={|g_0|^2\over N^2}{\partial\over\partial\beta}\sum_{l=1}^N\left({\beta\over 1-\beta \zeta^{j-l}}+\zeta^{l-j}\right). \notag
\end{align}
The first term in the third line is evaluated by using (59) whereas the second term becomes zero since  $\sum_{l=1}^N\zeta^{l-j}=\zeta^{1-j}(1-\zeta^{N})/(1-\zeta)=0$.
Consequently,
$$(CC^*)_{jj}={c\over N}{1+(N-1)\alpha^2\over 1-\alpha^2}. \eqno(61)$$
\par
Referring to (60) and (61), the characteristic polynomial  (43) takes the tractable form 
$$\chi_{CC^*}(\Lambda)=\left|\left(\left\{{c\over N}{1+(N-1)\alpha^2\over 1-\alpha^2}-\Lambda\right\}\delta_{jk}+{c\over N}(1-\delta_{jk})\right)_{1\leq j,k\leq N}\right|, $$
where $\delta_{jk}$ is Kronecker's delta. Indeed, the determinant can be evaluated in a sequence of steps, as  shown in the following.  After extracting the factor $c/N$ from the $j$th row for
$j=1,2, ..., N$, we add the $j$th column to the first column for $j=2,3, ..., N$ and them extract the factor ${N\over 1-\alpha^2}-{N\over c}\Lambda$ from the first column to obtain
$$\chi_{CC^*}(\Lambda)=\left({c\over N}\right)^N\left({N\over 1-\alpha^2}-{N\over c}\Lambda\right)\tilde \chi_{CC^*}(\Lambda), $$
with
$$\tilde \chi_{CC^*}(\Lambda)=\begin{vmatrix} 1    &1    & \cdots& 1 \\
                                      1  &\gamma+1 &\cdots&1\\
                                      \vdots& \vdots & \ddots &\vdots\\
                                      1& 1& \cdots& \gamma+1
                                      \end{vmatrix},\quad \gamma={N\alpha^2\over 1-\alpha^2}-{N\over c}\Lambda. $$
  Subtracting the first row from the $j$th row for $j=2,3, ..., N$, $\tilde \chi_{CC^*}(\Lambda)$ is transformed to the determinant of an upper triangular matrix
  and is evaluated simply to give $\tilde \chi_{CC^*}(\Lambda)=\gamma^{N-1}$. It turns out that
  $$ \chi_{CC^*}(\Lambda)=\left({c\over N}\right)^N\left({N\over 1-\alpha^2}-{N\over c}\Lambda\right)\left({N\alpha^2\over 1-\alpha^2}-{N\over c}\Lambda\right)^{N-1}. $$
Thus, the characteristic equation $\chi_{CC^*}(\Lambda)=0$ yields the multiple eigenvalues (57).  \hspace{\fill}$\Box$ \par
\par
\medskip
\noindent {\it Remark} 4. It follows from (52) with $|g_0|^2=c(1-\alpha^2)$ that
$$|C|=\prod_{j=1}^N\hat\lambda_j=\prod_{j=1}^N{g_0\alpha^{2(j-1)/N} \over 1-\alpha^2}, $$
and hence 
$$|CC^*|=\prod_{j=1}^N\Lambda_j=\prod_{j=1}^N(\hat\lambda_j\hat\lambda_j^*)=\prod_{j=1}^N{c\alpha^{4(j-1)/N} \over 1-\alpha^2}=\left({c\over 1-\alpha^2}\right)^N\alpha^{2(N-1)}.$$
The quantity $|CC^*|$ is also 
computed from (57) to give the same result. Actually, 
$$|CC^*|={c\over 1-\alpha^2}\left({c\alpha^2\over 1-\alpha^2}\right)^{N-1}=\left({c\over 1-\alpha^2}\right)^N\alpha^{2(N-1)}.$$
\par
\bigskip
\noindent {\bf 4 MULTISOLITON SOLUTIONS} \par
\medskip
\noindent{\bf 4.1 $N$-soliton solution} \par
\medskip
\noindent The goal of this section is to prove the following theorem. \par
\medskip
\noindent {\bf Theorem 3.} \ {\it The cubic Szeg\"o  equation (1) admits the $N$-soliton solution
$$u={\tilde g\over \tilde f},\quad |u|^2={\rm i}\,{\partial\over \partial t}\,{\rm ln}\,{\tilde f^*\over\tilde f}, \quad \tilde f=|\tilde F|, \quad \tilde g=|\tilde G|, \eqno(62a)$$
where $\tilde F=(\tilde f_{jk})_{1\leq j,k\leq N}$ is an $N\times N$ matrix with elements
$$\tilde f_{jk}=\left\{\kappa_j(x-v_jt-x_{j0})+{{\rm i}\over 2\lambda_j}\right\}\delta_{jk}
+{{\rm i}\over \lambda_j^2-\lambda_k^2}\left(\lambda_j{s_j\over s_k}-\lambda_k{s_k\over s_j}\right)(1-\delta_{jk}), \eqno(62b)$$
and $\tilde G=(\tilde g_{jk})_{1\leq j,k\leq N+1}$ is an $(N+1)\times (N+1)$ matrix given by
$$\tilde G=\tilde F((1/s_k);(1/s_j)),\eqno(62c)$$
with
$$\kappa_j={2\pi\over \lambda_j\nu_j^2}, \quad v_j={\lambda_j^2\nu_j^2\over 2\pi}, \quad s_j=e^{{{\rm i}\over 2}(\lambda_j^2t+\phi_{j0})},
\quad \nu_j, \ x_{j0}, \ \phi_{j0} \in \mathbb{R},\quad (j=1, 2, ..., N), \eqno(62d)$$
and the parameters $\lambda_j$ are imposed on the condition
$$\lambda_1>\lambda_2> ... >\lambda_N>0. \eqno(63) $$
\par
The tau-functions $\tilde f$ and $\tilde g$ satisfy the set of bilinear equations
$${\rm i}D_t{\tilde f}^*\cdot {\tilde f}-{\tilde g}^*{\tilde g}=0, \eqno(64)$$
$$D_t{\tilde f}^*\cdot {\tilde g}={\tilde h}^*{\tilde f}, \eqno(65)$$
where ${\tilde h}$ is the tau-function given by
$${\tilde h}=-{\rm i}|\tilde F((\lambda_ks_k);(\lambda_js_j))|. \eqno(66)$$}
\par
First, we derive the $N$-soliton solution by means of the long-wave limit of the $N$-phase solution. Subsequently, we perform
the direct proof of the former solution. In deriving (64) and (65) in Theorem 3, the invertibility of the matrix $\tilde F$ plays a central role. It  will be established later in Proposition 9. \par
\medskip 
\noindent {\bf Proposition 8.}\ {\it Let $k_j=\epsilon, \mu_j=\lambda_j-\epsilon \rho_j$ and $\theta_{j0}+\phi_{j0}=\epsilon x_{j0}$ for $j=1, 2, ..., N$
in the $N$-phase solution (12) with (13) and (14) and take the limit $\epsilon\rightarrow +0$. Then, the $N$-phase solution
reduces to the $N$-soliton solution (62).}\par
\medskip
\noindent{\it Proof.}\ In the limit $\epsilon\rightarrow\rightarrow +0$, the leading-order asymptotics of the parameters are found to be
$$\phi_j+\theta_j=2\epsilon\left(\lambda_j\rho_jt+{1\over 2}\,x_{j0}\right)+O(\epsilon^2),$$
$$\lambda_j^2-\mu_j^2=2\epsilon\lambda_j\rho_j+O(\epsilon^2).$$
The diagonal elements of the matrix $F$ from (13a) have the asymptotic form
$$f_{jj}=-{\rm i}\left\{{1\over 2\rho_j}(x-2\lambda_j\rho_j\,t-x_{j0})+{{\rm i}\over 2\lambda_j}\right\}+O(\epsilon).$$
If we put
$${1\over 2\rho_j}={2\pi\over\lambda_j\nu_j^2}=\kappa_j, \quad {\lambda_j^2\nu_j^2\over 2\pi}=v_j, \quad (j=1, 2, ..., N),$$
the  expression of $f_{jj}$ can be written in the form
$$f_{jj}=-{\rm i}\left\{\kappa_j(x-v_jt-x_{j0})+{{\rm i}\over 2\lambda_j}\right\}+O(\epsilon).$$
On the other hand, the non-diagonal elements reduce simply to
$$f_{jk}={1\over \lambda_j^2-\lambda_k^2}\left(\lambda_j{s_j\over s_k}-\lambda_k{s_k\over s_j}\right)+O(\epsilon), \quad (j\not=k).$$
Combining the above expressions of $f_{jj}$ and $f_{jk}$, the tau-function $f$ is found to have a limiting form $f=(-{\rm i})^N|\tilde F|$.
By means of the same procedure, the tau-functions $g$ and $h$ reduce respectively to $g=(-{\rm i})^{N+1}|\tilde F((1/s_k);(1/s_j))|$
and $h=-{\rm i}|\tilde F((\lambda_ks_k);(\lambda_js_j))|$. 
Last, since  $c=O(\epsilon)$, the bilinear equations (10) and (11) recast respectively to (64) and (65).  \hspace{\fill}$\Box$ \par
\par
\medskip
\noindent{\bf 4.2 Proof of Theorem 3}\par
\medskip
\noindent The proof of Theorem 3 is performed on the basis of some formulas  of determinants which are summarized in Lemma 3 below.\par
\medskip
\noindent {\bf Lemma 3.}\par
\medskip
$$\tilde f^*=|\tilde F|+{\rm i}|\tilde F((s_k/\lambda_k);(1/s_j))|, \eqno(67a)$$
$$\tilde f_t=|\tilde F((1/s_k);(\lambda_js_j))|, \eqno(67b)$$
$${\tilde f_t}^*=|\tilde F((s_k/\lambda_k);(\lambda_j^2/s_j))|, \eqno(67c)$$
$${\tilde g}^*=|\tilde F((s_k/\lambda_k);(\lambda_js_j))|, \eqno(67d)$$
$$\tilde g_t=-{\rm i}|\tilde F((1/s_k);(\lambda_j^2/s_j))|, \eqno(67e)$$
$$|\tilde F((1/s_k),(\lambda_ks_k);(1/s_j),(s_j/\lambda_j))|={\rm i}|\tilde F((\lambda_k^2/s_k);(s_j/\lambda_j))|-{\rm i}|\tilde F((1/s_k);(\lambda_js_j))|. \eqno(67f)$$
\medskip
\noindent The proof of Lemma 3 is given in Appendix 3. \par
\medskip
\noindent{\bf 4.2.1 Proof of (64)}\par
\medskip
\noindent Let $\tilde P={\rm i}D_t{\tilde f}^*\cdot {\tilde f}-{\tilde g}^*{\tilde g}$. Substituting (62b), (62c) and (67a)-(67d) into $\tilde P$ and applying Jacobi's formula, we deduce
$$\tilde P={\rm i}\Big\{|\tilde F((s_k/\lambda_k);(\lambda_j^2/s_j))|-|\tilde F((1/s_k);(\lambda_js_j))|+{\rm i}|\tilde F((1/s_k),(\lambda_ks_k);(1/s_j),(s_j/\lambda_j))|\Big\}|\tilde F|.$$
In view of (67f) and the fact that $\tilde F$ is a symmetric matrix, $\tilde P$ becomes zero identically.  \hspace{\fill}$\Box$ \par
\medskip
\noindent {\bf 4.2.2 Proof of (65)}\par
\medskip
\noindent Let $\tilde Q=D_t{\tilde f}^*\cdot {\tilde g}$.
We introduce (62c), (67a), (67c) and (67e) into $\tilde Q$ and then apply Jacobi's formula to the resulting expression to obtain
\begin{align}
\tilde Q &= \Big\{|\tilde F((s_k/\lambda_k),(1/s_k);(\lambda_j^2/s_j),(1/s_j))|+{\rm i}|\tilde F((1/s_k);(\lambda_j^2/s_j))|\Big\}|\tilde F| \notag \\
         &=\begin{vmatrix} \tilde F &\left({\lambda_j^2\over s_j}\right)^T &\left({1\over s_j}\right)^T \\
                           (s_k/\lambda_k) & 0 & -{\rm i} \\
                           (1/s_k) & 0 & 0 \end{vmatrix}|\tilde F| \notag 
\end{align}
We multiply $N+2$ column of the determinant by $-{\rm i}s_k/\lambda_k$, add it to the $k$th column respectively for $k=1, 2, ..., N$ and then use the formula
$$\tilde f_{jk}-{\rm i}\,{s_k\over \lambda_ks_j}={\lambda_j\over \lambda_k}\, {\tilde f_{jk}}^*.$$
We modify the resulting expression of $\tilde Q$ as
\begin{align}
\tilde Q &={\rm i}\begin{vmatrix} \left({\lambda_j\over \lambda_k}\, {\tilde f_{jk}}^*\right) &\left({\lambda_j^2\over s_j}\right)^T  \\
                           (1/s_k) & 0 \end{vmatrix}|\tilde F| \notag  \\
         &={\rm i}\begin{vmatrix} \tilde F^*&\left({\lambda_j/ s_j}\right)^T  \\
                           (\lambda_k/s_k) & 0 \end{vmatrix}|\tilde F| \notag \\
         &={\tilde h}^*\tilde f, \notag
\end{align}
where the relations $\tilde F={\tilde F}^T$ and $s_j^*=1/s_j$ have been use in passing to the third line.
Consequently, $\tilde Q={\tilde h}^*\tilde f$, which completes the proof of (65). \hspace{\fill}$\Box$ \par
\medskip
\noindent {\it Remark} 5. \ The $N$-soliton solution (62) has been obtained for the first time by Pocovnicu$^{2}$ whereby its derivation is based on
the spectral analysis of the Hankel operator.  Recall that the Hankel operator plays the central role 
in the Lax pair structure of Equation (1). The method developed in our paper is substantially different from that used by Pocovnicu in several points.
First, it does not rely on the complete integrability of the Equation (1) and hence the knowledge of the IST is not required. 
Second, the proof of the $N$-soliton solution is done by means of an elementary theory of determinants.\par 
\medskip
\noindent{\bf 4.3 Invertibility of the matrix $\tilde F$}\par
\medskip
\noindent The invertibility of the matrix $\tilde F$ is a key point in deriving the bilinear equations (64) and (65).
While the invertibility of the matrix $F$ from (13a) would survive after the long-wave limit has been taken, it is not an obvious issue.
Hence, we present its direct proof in the following proposition. \par
\medskip
\noindent {\bf Proposition 9.}\ {\it The matrix $\tilde F$ is invertible, or equivalently $|\tilde F|\not=0$ in the upper-half
complex plane. } \par
\medskip
\noindent{\it Proof.}\ Assume that $|\tilde F|=0$ occurs at $x_p={\rm Re}\,x_p+{\rm i}\,{\rm Im}\,x_p$. Then, there exists  a nonzero vector $(\Psi_j)_{1\leq j\leq N}$
satisfying  the system of linear algebraic equations
$$\sum_{k=1}^N\tilde f_{jk}\Psi_k=0, \quad (j=1, 2, ..., N),\quad \sum_{j=1}^N|\Psi_j|^2\not=0. $$
Multiplying this equation by $\Psi_j^*$ and adding with respect to $j$, one has $\sum_{j,k=1}^N\Psi_j^*\tilde f_{jk}\Psi_k=0$.
It follows from the imaginary part of this equation that
$${\rm Im}\,x_p\sum_{j=1}^N\kappa_j|\Psi_j|^2+{1\over 2}\sum_{j,k=1}^N{{s_j\over s_k}+{s_k\over s_j}\over \lambda_j+\lambda_k}\,\Psi_j^*\Psi_k=0. $$
This expression can be modified by noting the identity $(\lambda_j+ \lambda_k)^{-1}=\int_0^\infty e^{-(\lambda_j+ \lambda_k)s}\,ds$ and the relation $s_j^*=1/s_j$  to give
$${\rm Im}\,x_p\sum_{j=1}^N\kappa_j|\Psi_j|^2
+{1\over 2}\,\sum_{j,k=1}^N\int_0^\infty e^{-(\lambda_j+ \lambda_k)s}\left\{\left({\Psi_j\over s_j}\right)^*\left({\Psi_k\over s_k}\right)+(s_j\Psi_j)^*(s_k\Psi_k)\right\}ds=0. $$
After a few manipulations, one finds that
$${\rm Im}\,x_p=-{1\over 2\sum_{j=1}^N\kappa_j|\Psi_j|^2}\int_0^\infty\left\{\left|\sum_{j=1}^Ne^{-\lambda_js}{\Psi_j\over s_j}\right|^2+\left|\sum_{j=1}^Ne^{-\lambda_js}{s_j\Psi_j}\right|^2\right\}ds.$$
Since $\sum_{j=1}^N\kappa_j|\Psi_j|^2>0$, the above relation implies that ${\rm Im}\,x_p<0$ and hence $|\tilde F|\not=0$ in ${\rm Im}\, x\geq 0$. 
This completes the proof of Proposition 8. \hspace{\fill}$\Box$ \par
\medskip
\noindent{\bf 4.4 Asymptotic behavior of the $N$-soliton solution}\par
\medskip
\noindent We investigate the asymptotic behavior of the $N$-soliton solution for large time and see a remarkable 
feature of the soliton interaction. \par
 \par
\medskip
\noindent {\bf Proposition 10.}\ {\it The asymptotic form of the $N$-soliton solution (62) for large time is represented by a superposition of the
one-soliton solutions. To be more specific, it reads
$$u\sim -\sum_{j=1}^N{{e^{-{\rm i}(\lambda_j^2t+\phi_{j0})}\over \kappa_j(x-v_jt-x_{j0})+{{\rm i}\over 2\lambda_j}}}, \quad t\rightarrow \pm\infty, \eqno(68)$$
showing that the solitons exhibit no phase shift after collisions between them.}
\par
\medskip
\noindent{\it Proof.}\ Let $\theta_j=x-v_jt-x_{j0}$. Take the limit $t\rightarrow -\infty$ with $\theta_j$ being fixed. 
The definition of the velocity from (62d) with the ordering of $\lambda_j$ from (62d) implies that $v_j>v_k$ for $j<k$.
Taking into account this inequality, we see in the above limit that $\theta_k\rightarrow -\infty$ for $k=1, 2, ..., j-1$ and
$\theta_k\rightarrow +\infty$ for $k=j+1,j+ 2, ..., N$, respectively. In this setting, the tau-functions $\tilde f$ and $\tilde g$ from (62)
have the leading-order asymptotics
$$\tilde f\sim \left(\kappa_j\theta_j+{{\rm i}\over 2\lambda_j}\right)\prod_{\substack{k=1 \\(k\not=j)}}^N\kappa_k\theta_k, 
\quad \tilde g\sim -s_j^{-2}\prod_{\substack{k=1 \\(k\not=j)}}^N\kappa_k\theta_k.$$
Consequently, if one observes the collision process in the coordinate system at rest,  the asymptotic form of $u=\tilde g/\tilde f$ follows immediately as indicated by (68). 
In the limit $t \rightarrow +\infty$,
on the other hand, the same limiting forms of the tau-functions are obtained, leading to the asymptotic form (68).\hspace{\fill}$\Box$ \par
\medskip
\medskip
\noindent {\it Remark} 6. \ The asymptotic expression (68) has been derived in analyzing an explicit formula for the $N$-soliton solution obtained in Ref. 2.
We recall that the similar feature of the solution has been found for the first time in  the interaction process of the algebraic (or rational) solitons
of the Benjamin-Ono equation.$^{22-24}$  
\par
\bigskip
\noindent {\bf 5 CONCLUDING REMARKS} \par
\medskip
\noindent In this paper, we have developed a systematic method for proving the multiphase and multisoliton solutions of the cubic Szeg\"o equation.
The bilinearization of the equation is the starting point in our analysis. The subsequent  proof of the solutions proceeds in a straightforward way.
We have also presented an alternative proof based on the Lax pair (2). Specifically, it was shown by an elementary computation that both the spatial and temporal parts of the
Lax pair are satisfied by the eigenfunctions for the $N$-phase solution. In addition, we have addressed the eigenvalue problem associated with a special class of periodic solutions whereby
we have obtained   the analytical expressions of the eigenvalues. \par
\bigskip
 \noindent {\bf APPENDIX A  PROOF OF LEMMA 1} \par
  \bigskip
  \noindent First, we enumerate the basic formulas of  determinants which are used frequently in the   proof.   Among them, Jacobi's identity
  will play an important role. See, for example Ref. 25. \par 
  $${\partial |F|\over \partial x}=\sum_{j,k=1}^N{\partial f_{jk}\over \partial x}\,F_{jk}, \eqno(A.1)$$
$$\begin{vmatrix} F & {\bf a}^T\\ {\bf b} & z\end{vmatrix}\equiv |F((b_k);(a_j))|=|F|z-\sum_{j,k=1}^NF_{jk}a_jb_k,  \eqno(A.2)$$
$$|F((a_k); (c_j))||F((b_k); (d_j))|-|F((a_k); (d_j))||F((b_k); (c_j))|$$
$$=|F((a_k), (b_k); (c_j), (d_j))||F|,\ ({\rm Jacobi's\ identity}), \eqno(A.3)$$
$$\delta_{jk}|F|=\sum_{l=1}^Nf_{jl}F_{kl}=\sum_{l=1}^Nf_{lj}F_{lk}, \eqno(A.4)$$
\begin{align}
F_{jk} &= \sum_{l=1}^Nf_{lm}\,F_{jl, km}\quad (k\not=m) \tag{A.5a} \\
&= \sum_{m=1}^Nf_{lm}\,F_{jl, km}\quad (j\not=l), \tag{A.5b}
\end{align}
$$|F((a_k), (b_k); (c_j), (d_j))|=\sum_{\substack{j,k,l,m=1\\(j\not=k,l\not=m)}}^Na_lb_mc_jd_kF_{jk,lm}. \eqno(A.6)$$
 \bigskip
 \bigskip
\noindent{\bf A.1 Proof of (15a)}\par
  \medskip
  \noindent Using the relation 
  $$f_{jk}^*=-{\lambda_j\over \mu_kz_k}\,e^{{\rm i}(\phi_j+\theta_k)}\left(f_{jk}-{1\over \lambda_j}\right), \eqno(A.7)$$
  which follows from (13a), and then  extracting the factors $\lambda_je^{{\rm i}\phi_j}$ from the $j$th row and $-e^{{\rm i}\theta_k}/(\mu_kz_k)$  from the $k$th column for
  $j, k=1, 2, ..., N$, respectively, one obtains
   $$f^* =\kappa\left|\left(f_{jk}-{1\over \lambda_j}\right)\right| 
      =\kappa\begin{vmatrix} f_{jk} & \left({1\over \lambda_j}\right)^T  \\
                              (1)& 1\end{vmatrix} \notag \\
      =\kappa(|F|+|F((1);(1/\lambda_j))|).   $$   
  \medskip 
  \noindent{\bf A.2 Proof of (15b)}\par
  \medskip
  \noindent Referring to (A.1) and (A.2)
  \begin{align}
      f_t&={\rm i}\sum_{j,k=1}^N\mu_kz_ke^{-{\rm i}(\phi_j+\theta_k)}F_{jk} \notag \\
         &=-{\rm i}|F((\mu_kz_ke^{-{\rm i}\theta_k});(e^{-{\rm i}\phi_j}))|. \notag
  \end{align}
      \medskip 
  \noindent{\bf A.3 Proof of (15c)}\par
  \medskip
  \noindent The complex conjugate expression of $f_t$ from (15b) is given by 
      $$f_t^*={\rm i}\begin{vmatrix} (f_{jk}^*) & \left(e^{{\rm i}\phi_j}\right)^T  \\
                              (\mu_ke^{{\rm i}\theta_k}/z_k)& 0\end{vmatrix}. $$
  Substituting $f_{jk}^*$ from (A.7) and extracting the factors $\lambda_je^{{\rm i}\phi_j}$ from the $j$th row and $-e^{{\rm i}\theta_k}/(\mu_kz_k)$  from the $k$th column for
  $j, k=1, 2, ..., N$, respectively, the above expression becomes
  $$f_{jk}^*={\rm i}\kappa\begin{vmatrix} \left(f_{jk}-{1\over \lambda_j}\right) & \left({1\over \lambda_j}\right)^T  \\
                              (-\mu_k^2)& 0\end{vmatrix}
                              =-{\rm i}\kappa|F((\mu_k^2);(1/\lambda_j))|. $$
  \medskip 
  \noindent{\bf A.4 Proof of (15d)}\par
  \medskip
  \noindent The expression of $g$ follows simply from (13b) and the definition (5).
  \par  
  \medskip
  \noindent{\bf A.5 Proof of (15e)}\par
  \medskip
  \noindent  If one takes the complex conjugate of $g$ from (15d), uses (A.7) and  extracts the factors $\lambda_je^{{\rm i}\phi_j}$ from 
  the $j$th row and $-e^{{\rm i}\theta_k}/(\mu_kz_k)$  from the $k$th column for   $j, k=1, 2, ..., N$, respectively,
     one can deduce it to
      $$g^*=-\begin{vmatrix} (f_{jk}^*) & \left(e^{{\rm i}\phi_j}\right)^T  \\
                              (1)& 0\end{vmatrix}
      =\kappa\begin{vmatrix} \left(f_{jk}-{1\over \lambda_j}\right) & \left({1\over \lambda_j}\right)^T  \\
                              (\mu_kz_ke^{-{\rm i}\theta_k})& 0\end{vmatrix}
      =\kappa|F((\mu_kz_ke^{-{\rm i}\theta_k});(1/\lambda_j))|. $$
  \medskip 
  \noindent{\bf A.6 Proof of (15f)}\par
  \medskip
  \noindent    Applying (A.1) to (15d), one obtains
      \begin{align}
      g_t&=-|F((1);(-{\rm i}\lambda_j^2e^{-{\rm i}\phi_j}))|-|F((1),(\mu_kz_ke^{-{\rm i}\theta_k});(e^{-{\rm i}\phi_j}),(e^{-{\rm i}\phi_j}))| \notag \\
         &={\rm i}|F((1);(\lambda_j^2e^{-{\rm i}\phi_j}))|, \notag
         \end{align}
      where the second term in the first line vanishes identically since the last two columns  of the determinant coincide. 
      \par
   \medskip 
   \noindent{\bf A.7 Proof of (15g)}\par
  \medskip
      \noindent If one puts $a_k=1, b_k=\mu_kz_ke^{-{\rm i}\theta_k}, c_j=e^{-{\rm i}\phi_j}, d_j=1/\lambda_j$ in (A.6) and introduces the matrix element $f_{jk}$ from (13a), one has
  \begin{align}
      L &\equiv \left|F((1),(\mu_kz_ke^{-{\rm i}\theta_k});(e^{-{\rm i}\phi_j}),(1/\lambda_k))\right| \notag \\
        &=\sum_{\substack{j,k,l,m=1\\(j\not=k,l\not=m)}}^N\mu_mz_me^{-{\rm i}(\theta_m+\phi_j)}{1\over \lambda_k}F_{jk, lm} \notag \\
      &=\sum_{\substack{j,k,l,m=1\\(j\not=k,l\not=m)}}^N\big\{\lambda_j-(\lambda_j^2-\mu_m^2)f_{jm}\big\}{1\over \lambda_k}F_{jk, lm}. \notag
      \end{align}
  In view of the relation $F_{jk,lm}=-F_{kj,lm}$ which comes from the definition (6) and (A.5), $L$ recasts to
\begin{align}
    L &={1\over 2}\sum_{\substack{j,k,l,m=1\\(j\not=k,l\not=m)}}^N{\lambda_j\over \lambda_k}(F_{jk,lm}+F_{jk,ml})
             +\sum_{\substack{j,k,l,m=1\\(j\not=k,l\not=m)}}^N{\lambda_j^2\over \lambda_k}f_{jm}F_{kj,lm}
             -\sum_{\substack{j,k,l,m=1\\(j\not=k,l\not=m)}}^N{\mu_m^2\over \lambda_k}f_{jm}F_{kj,lm}\notag \\
             &=\sum_{\substack{j,k,l=1 \\ (j\not=k)}}^N{\lambda_j^2\over \lambda_k}F_{kl}-\sum_{\substack{k,l,m=1 \\ (l\not=m)}}^N{\mu_m^2\over \lambda_k}F_{kl} \notag \\
             &=\sum_{j,k,l=1}^N{\lambda_j^2\over \lambda_k}F_{kl}-\sum_{k,l=1}^N\lambda_kF_{kl}-\sum_{k,l,m=1}^N{\mu_m^2\over \lambda_k}F_{kl}+\sum_{k,m=1}^N{\mu_m^2\over \lambda_k}F_{km} \notag \\
             &=-\sum_{j=1}^N(\lambda_j^2-\mu_j^2)|F((1);(1/\lambda_j))|+|F((1);(\lambda_j))|-|F((\mu_k^2);(1/\lambda_j))|. \notag
  \end{align}
      \medskip 
  \noindent{\bf A.8 Proof of (15h)}\par
  \medskip
       \noindent If one puts $a_k=1, b_k=\mu_kz_ke^{-{\rm i}\theta_k}, c_j=e^{-{\rm i}\phi_j}, d_j=e^{-{\rm i}\phi_j}$ in (A.6), the left-hand of the determinant 
       vanishes identically since the $N+1$ and $N+2$ columns coincide.
       The right-hand side can be evaluated along the same lines as used in A.7. 
       \par
       \bigskip
   \noindent {\bf APPENDIX B  PROOF OF LEMMA 2} \par
  \bigskip
  \noindent{\bf B.1 Proof of (27a)}\par
  \medskip
  \noindent Taking the complex conjugate of (25) and using (A.7) with $\hat f_{jk}=f_{jk}-1/\lambda_j$, one deduces
  \begin{align} {\hat g}^* &=-\begin{vmatrix} F^* & (e^{{\rm i}\phi_j}s^{j-1})^T \\
                               (1) & 0 \end{vmatrix} \notag \\
                           &=\prod_{j=1}^N{\lambda_je^{{\rm i}(\phi_j+\theta _j)}\over -\mu_jz_j}\begin{vmatrix} \hat F &\left({s^{j-1}\over \lambda_j}\right)^T \\
                               (\mu_kz_ke^{-{\rm i}\theta_k}) & 0 \end{vmatrix} \notag \\
                           &=\kappa|\hat F((\mu_kz_ke^{-{\rm i}\theta_k}); (s^{j-1}/\lambda_j))|.\notag
  \end{align}
  \medskip
  \noindent{\bf B.2 Proof of (27b)}\par
  \medskip
  \noindent Referring to (23) 
  $$\sum_{j=1}^N\lambda_je^{{\rm i}\phi_j}g_js^{j-1}=\sum_{j, k=1}^N\lambda_js^{j-1}F_{jk}=-|F((1); (\lambda_js^{j-1}))|=-|\hat F((1); (\lambda_js^{j-1}))|,$$
  where the last line follows simply by multiplying the $(N+1)$th row of the previous determinant by $1/\lambda_j$ and subtracting it from the $j$th rows
  for $j=1, 2, ..., N$.
      \par
      \medskip
  \noindent{\bf B.3 Proof of (27c)}\par
  \medskip
   \noindent   By the same procedure as used in B.1,  the expression of $f^*$  follows from (15a).
           \par
      \medskip
  \noindent{\bf B.4 Proof of (27d)}\par
  \medskip
  \noindent The left-hand side of (27d) which is denoted by $L$ is expressed by employing Formula (A.2) as
  $$L=-\sum_{j,k=1}^N\mu_kz_ke^{-{\rm i}(\phi_j+\theta _k)}\hat F_{jk}=-\sum_{j,k=1}^N\left\{{\mu_k^2\over \lambda_j}-(\lambda_j^2-\mu_k^2)\hat f_{jk}\right\}\hat F_{jk}, $$
            where,  in the last line, the definition of $\hat f_{jk}$ from (27a) has been used. In view of (A.2) and (A.4), $L$ becomes
            $$L=|\hat F((\mu_k^2); (1/\lambda_j))|+\sum_{j=1}^N(\lambda_j^2-\mu_j^2)|\hat F|.$$
      \par
      \medskip
  \noindent{\bf B.5 Proof of (27e)}\par
  \medskip
      \noindent Let the left-hand side of (27e) be $L$. As seen easily, the matrix $\hat F$ in $L$ can be replaced by $F$. 
      By developing the same procedure as used in proving (15g), one obtains
      $$L=-\sum_{j=1}^N(\lambda_j^2-\mu_j^2)| F((1);(s^{j-1}/\lambda_j))|+| F((1);(\lambda_js^{j-1}))|-|F((\mu_k^2);(s^{j-1}/\lambda_j))|. $$
      The matrix $ F$ of the first two terms can be replaced by $\hat F$, which gives (27e). 
      \par
      \medskip
  \noindent{\bf B.6 Proof of (27f)}\par
  \medskip
      \noindent Referring to the definition of $\hat F$ from (27a)
      $$|\hat F|=\begin{vmatrix} \left(f_{jk}-{1\over \lambda_j}\right) & \left({1\over \lambda_j}\right)^T \\
                               (0) & 1 \end{vmatrix} 
                               =\begin{vmatrix} F & \left({1\over \lambda_j}\right)^T \\
                               (1) & 1 \end{vmatrix}
                               =|F|+\begin{vmatrix} F & \left({1\over \lambda_j}\right)^T \\
                               (0) & 1 \end{vmatrix}.$$
    One can replace the matrix  $F$ in the second term of the last line  by $\hat F$ to obtain (27f).
      \par
            \bigskip
   \noindent {\bf APPENDIX C  PROOF OF LEMMA 3} \par
  \bigskip
      \noindent{\bf C.1 Proof of (67a)}\par
  \medskip
      \noindent  It follows from (62b) that
      $$\tilde f_{jk}^*={\lambda_k\over \lambda_j}\,\tilde f_{jk}-{\rm i}\,{s_k\over \lambda_js_j}. \eqno(C.1)$$
      One uses this relation to obtain
      $$ \tilde f^*=\left|\left({\lambda_k\over \lambda_j}\,\tilde f_{jk}-{\rm i}\,{s_k\over \lambda_js_j}\right)\right|
      =\begin{vmatrix} \left({\lambda_k\over \lambda_j}\,\tilde f_{jk}\right) & {\rm i}\left({1\over \lambda_js_j}\right)^T \\
                               (s_k) & 1 \end{vmatrix}
      =\begin{vmatrix} \left(\tilde f_{jk}\right) & {\rm i}\left({1\over s_j}\right)^T \\
                               \left({s_k\over \lambda_k}\right) & 1 \end{vmatrix}, $$
      where,  in passing to the last line, one subtracted the factor $1/\lambda_j$ from the $j$th row and the factor $\lambda_k$
      from the $k$th column respectively for $j, k=1, 2, ..., N$. An elementary algebra with the aid of the property of the
      determinant yields (67a). \par
      \medskip
      \noindent{\bf C.2 Proof of (67b)}\par
  \medskip
      \noindent  Applying the differential rule (A.1) to $\tilde f$ from (62a) with (62b) gives
      \begin{align} 
      \tilde f_t &=-\sum_{j=1}^N\lambda_j\tilde F_{jj}-{1\over 2}\sum_{\substack{j,k=1\\ (j\not=k)}}^N\left(\lambda_j{s_j\over s_k}+\lambda_k{s_k\over s_j}\right)\tilde F_{jk} \notag \\
      &= -{1\over 2}\sum_{j,k=1}^N\left(\lambda_j{s_j\over s_k}+\lambda_k{s_k\over s_j}\right)\tilde F_{jk} \notag \\
      &=-\sum_{j,k=1}^N\lambda_j{s_j\over s_k}\tilde F_{jk},\notag
      \end{align}
      where, in passing to the last line, the relation $\tilde F_{jk}=\tilde F_{kj}$ has been used which comes from the property $\tilde F=\tilde F^T$.
      Referring to (A.2), the above expression is written as (67b). \par
      \medskip
      \noindent{\bf C.3 Proof of (67c)}\par
  \medskip
      \noindent It follows from the complex conjugate of (67b), (C.1) and the relation $s_j^*=1/s_j$ that
      $$\tilde f_t^*=\begin{vmatrix} \left({\lambda_k\over \lambda_j}\,\tilde f_{jk}-{\rm i}\,{s_k\over \lambda_js_j} \right) & \left({\lambda_j\over s_j}\right)^T \\
                               (s_k) & 0 \end{vmatrix}
                               =\begin{vmatrix} \left({\lambda_k\over \lambda_j}\,\tilde f_{jk} \right) & \left({\lambda_j\over s_j}\right)^T \\
                               (s_k) & 0 \end{vmatrix}
                               =\begin{vmatrix}\tilde F & \left({\lambda_j^2\over s_j}\right)^T \\
                               ({s_k\over \lambda_k}) & 0 \end{vmatrix}. $$
      The second line follows if one multiplies ${\rm i}/(\lambda_js_j)$ by $(N+1)$th row and then adds it to the $j$th row for $j=1, 2, ..., N$ whereas
      the last line comes by multiplying $\lambda_j$ to the $j$th row and $1/\lambda_k$ by the $k$th column respectively for $j, k=1, 2, ..., N$.
      \par
       \medskip
      \noindent{\bf C.4 Proof of (67d)}\par
  \medskip
      \noindent One starts  from the complex conjugate of (62c). Use (C.1)  and the relation $s_j^*=1/s_j$ to obtain
      $$\tilde g^*=\begin{vmatrix} \left({\lambda_k\over \lambda_j}\,\tilde f_{jk}-{\rm i}\,{s_k\over \lambda_js_j} \right) & (s_j)^T \\
                               (s_k) & 0 \end{vmatrix}.$$
      Repeating  the same argument as already employed in C.2 to derive (67c) yields (67d). 
      \par
       \medskip
      \noindent{\bf C.5 Proof of (67e)}\par
  \medskip
      \noindent Applying the  differential rule (A.1) to $\tilde g=|\tilde G|$, one has
      $$\tilde g_t=\begin{vmatrix} \tilde F & \left({1\over s_j}\right)^T &  (\lambda_js_j)^T\\
                               ({1\over s_k}) & 0 &0 \\
                               ({1\over s_k}) & 0 &0 \end{vmatrix}
                               +\begin{vmatrix} \tilde F & -{{\rm }i\over 2}\left({\lambda_j^2\over s_j}\right)^T\\
                                                             ({1\over s_k}) & 0  \end{vmatrix}
                               +\begin{vmatrix} \tilde F &({1\over s_j} )^T\\
                                                            -{{\rm }i\over 2}\left({\lambda_j^2\over s_k}\right)  & 0  \end{vmatrix}. $$
      The first term vanishes identically since the $(N+1)$th row and the $(N+2)$th row coincide whereas the second term is equal to the third term
      due to the property $\tilde F=\tilde F^T$. Extracting the factor $-{\rm i}/2$, $\tilde g_t$ reduces to (67e). 
      \par
       \medskip
      \noindent{\bf C.6 Proof of (67f)}\par
  \medskip
      \noindent Let the left-hand side of (67f) be L. If one puts $a_k=1/s_k, b_k=\lambda_ks_k, c_j=1/s_j, d_j=s_j/\lambda_j$, $L$ becomes
      $$L=\sum_{\substack{j,k,l,m=1\\(j\not=k,l\not=m)}}^N{1\over s_j}{s_k\over\lambda_k}{1\over s_l}\lambda_ms_m\tilde F_{jk,lm}. $$
      It follows from (62b) that 
      $${\lambda_ms_m\over s_j}={\lambda_js_j\over s_m}-{\rm i}(\lambda_m^2-\lambda_j^2)\tilde f_{mj},$$
      and $\tilde f_{mj}=\tilde f_{jm}$. Taking into account these formulas as well as the definition $\tilde F_{jk,lm}=-\tilde F_{kj,lm}$,  $L$ recasts to
      \begin{align}
      L&= \sum_{\substack{j,k,l,m=1\\(j\not=k,l\not=m)}}^N{\lambda_js_j\over s_m}{s_k\over \lambda_ks_l}\tilde F_{jk, lm}
      +{\rm i}\sum_{\substack{j,k,l,m=1\\(j\not=k,l\not=m)}}^N{s_k\over \lambda_ks_l}(\lambda_m^2-\lambda_j^2)\tilde f_{jm}\tilde F_{kj,lm} \notag \\
      &=\sum_{\substack{j,k,l,m=1\\(j\not=k,l\not=m)}}^N{\lambda_js_j\over s_m}{s_k\over \lambda_ks_l}\tilde F_{jk, lm}
      +{\rm i}\left\{\sum_{\substack{k,l,m=1\\(l\not=m)}}^N{s_k\over \lambda_ks_l}\lambda_m^2\tilde F_{kl}
      -\sum_{\substack{j,k,l=1\\(j\not=k)}}^N{s_k\over \lambda_ks_l}\lambda_j^2\tilde F_{kl}\right\} \notag \\
      &={1\over 2}\sum_{\substack{j,k,l,m=1\\(j\not=k,l\not=m)}}^N{\lambda_js_j\over s_m}{s_k\over \lambda_ks_l}(\tilde F_{jk, lm}+\tilde F_{jk, ml})
       -{\rm i}\sum_{k,m=1}^N{s_k\over \lambda_ks_m}\lambda_m^2\tilde F_{km}+{\rm i}\sum_{k,l=1}^N{\lambda_ks_k\over s_l}\tilde F_{kl},\notag
       \end{align}
      where, in passing to the second line, Formula (A.5) has been used. The first term of the third line vanishes due to the relation $\tilde F_{jk,lm}=-\tilde F_{jk,ml}$.
      The remaining terms are modified by applying (A.2) and using the notation (5), giving (67f). 
      \par
      \bigskip
      \noindent{\bf ACKNOWLEDGMENT} \par
      \medskip
      \noindent This paper has been written in memory of Prof. David Kaup. I would like to express my gratitude to the editors of this special issue.
      My special thanks are due to  Prof. Taras Lakoba for inviting me to contribute to the issue. \par
      \bigskip
      \noindent{\bf DATA AVAILABILITY STATEMENT}\par
      \bigskip
      \noindent Data sharing is not applicable to this paper as no new data were created or analyzed in this study. \par

\newpage
 
\leftline{\bf REFERENCES}\par
\begin{enumerate}[{1.}]
\item G\'erard P, Grellier S. The cubic Szeg\"o equation. {\it Ann Sci Ecole Norm Super.} 2010; 43: 761-810.
\item Pocovnicu O. Explicit formula for the solution of the Szeg\"o equation on the real line and applications. {\it Discrete  Cont Dyn Syst.} 2011;31: 607-649.
\item Pocovnicu O. Traveling waves for the cubic Szeg\"o equation on the real line. {\it Anal PDE.} 2011; 4:379-404.
\item G\'erard P,  Grellier S. Invariant tori for the cubic Szeg\"o equation {\it Invent Math.} 2012; 187: 707-754
\item G\'erard P, Grellier S. An explicit formula for the cubic Szeg\"o equation. {\it Trans Amer Math Soc.} 2015; 367: 2979-2995.
\item G\'erard P, Grellier S. The cubic Szeg\"o equation and Hankel operators. {\it Asterisque.} 2017; 2017: 1-122.
\item G\'erard P  and Pushnitski A. Inverse spectral theory for a class of non-compact Hankel operators. {\it Mathematika.} 2019; 65: 132-156.
\item G\'erard P, Pushnitski.  The cubic Szeg\"o equation on the real line: Explicit formula 
and well-posedness on the Hardy class. {\it Comm. Math. Phys.} 2024; 405: 167.
\item Gardner CS, Greene JM, Kruskal MD, Miura R.  Method for solving the Korteweg-de Vries equation. {\it Phys Rev Lett.} 1967; 19: 1095-1097.
\item Ablowitz MJ, Segur H. {\it Solitons and the Inverse Scattering Transform.} SIAM;1981.
\item Dodd RK, Eilbeck JC, Gibbon JD, Morris HC. {\it Solitons and Nonlinear Wave Equations.} Academic; 1982.
\item Faddeev LD and Takhtajan LA.  {\it Hamiltonian Methods in the Theory of Solitons.} Springer; 2007.
\item Wahlquist H, Estabrook FB.  B\"acklund Transformation for solutions of the Korteweg-de Vries equation. {\it Phys Rev Lett.} 1973; 31: 1386-1390.
\item Dold A, Eckman B, eds. {\it B\"acklund Transformations (Lecture Notes in Math. 515).}  Springer; 1974.
\item Rogers C, Shadwick WF. {\it B\"acklund Transformations and Their Applications.} Academic; 1982.
\item Hirota R. Exact solution of the Korteweg-de Vries equation for multiple collisions of solitons. {\it Phys Rev Lett}. 1971; 27: 1192-1194.
\item Matsuno Y. {\it Bilinear Transformation Method.}  Academic; 1984.
\item Hirota R.  {\it The Direct Method in Soliton Theory.}  Cambridge University Press; 2004.
\item Matsuno Y.  Multiphase solutions and their reductions for a nonlocal nonlinear Schr\"odinger equation with focusing nonlinearity. {\it Stud Appl Math.} 2023; 151: 883-922.
\item G\'erard P, Lenzmann E. The Calogero-Moser derivative nonlinear Schr\"odinger equation {\it Comm. Pure Appl. Math.} 2024; 77: 4008-4062.
\item Gray RM. Toeplitz and circulant matrices: A review. {\it Foundations and trends in communications and information theory.} 2006; 2:155-239.
\item Matsuno Y.   Exact multi-soliton solution of the Benjamin-Ono equation. {\it J Phys A: Math Gen.} 1979; 12: 619-21.
\item Matsuno Y.  Interaction of the Benjamin-Ono solitons {\it J Phys A: Math Gen.} 1980; 13: 1519-1536.
\item Matsuno Y. 1995 Dynamics of interacting algebraic solitons {\it Int National J Mod Phys B.} 1995; 9: 1985-2081.
\item Vein R, Dale P. {\it Determinants and Their Applications in Mathematical Physics.}  Springer; 1999.

\end{enumerate}
\end{document}